\documentclass[12pt,reqno]{amsart}
\usepackage{amsmath}
\usepackage[latin1]{inputenc}
\usepackage{amssymb}
\usepackage{amscd}
\usepackage{mathrsfs}
\usepackage[pdftex]{graphicx}
\usepackage{graphicx}

\newcommand{\One}[0]{\ensuremath{\mathbf{1}}}

\newcommand{\E}[0]{\ensuremath{\mathbf{E}}}
\newcommand{\F}[0]{\ensuremath{\mathcal{F}}}

\newcommand{\Pe}[0]{\ensuremath{\mathbf{P}}}
\newcommand{\R}[0]{\ensuremath{\mathbb{R}}}
\newcommand{\Z}[0]{\ensuremath{\mathbb{Z}}}

\newtheorem{theo}{\sc{Theorem}}[section]
\newtheorem{lemm}[theo]{\sc{Lemma}}
\newtheorem{defi}[theo]{\sc{Definition}}

\newtheorem{cor}[theo]{\sc{Corollary}}
\newtheorem{rmq}[theo]{\sc{Remark}}

\begin{document}

\title[Turbulence in White-Forced Generalised Burgers Equation.]{Sharp Estimates for Turbulence in White-Forced Generalised Burgers Equation.}

\author{Alexandre Boritchev}

\date{}

\maketitle

\begin{center}
Centre de Mathematiques Laurent Schwartz,
\\
Ecole Polytechnique, Route de Saclay
\\
91128 Palaiseau Cedex, France.
\\
E-mail: alexandre.boritchev@gmail.com
\\
Telephone number: (+33) 1 69 33 49 19 
\\
Fax number: (+33) 1 69 33 49 49
\end{center}

\begin{abstract}
We consider the non-homogeneous generalised Burgers equation
\begin{equation} \nonumber
\frac{\partial u}{\partial t} + f'(u)\frac{\partial u}{\partial x} - \nu \frac{\partial^2 u}{\partial x^2} = \eta,\ t \geq 0,\ x \in S^1.
\end{equation}
Here $f$ is strongly convex and satisfies a growth condition, $\nu$ is small and positive, while $\eta$ is a random forcing term, smooth in space and white in time.
\\ \indent
For any solution $u$ of this equation we consider the quasi-
\\
stationary regime, corresponding to $t \geq T_1$, where $T_1$ depends only on $f$ and on the distribution of $\eta$. We obtain sharp upper and lower bounds for Sobolev norms of $u$ averaged in time and in ensemble. 
These results yield sharp upper and lower bounds for natural analogues of quantities characterising the hydrodynamical turbulence. All our bounds do not depend on the initial condition or on $t$ for $t \geq T_1$, and hold uniformly in $\nu$.
\\ \indent
Estimates similar to some of our results have been obtained by Aurell, Frisch, Lutsko and Vergassola on a physical level of rigour; we use an argument from their article.
\end{abstract}

\maketitle

\tableofcontents

\section*{Introduction}

\indent
The generalised one-dimensional space-periodic Burgers equation
\begin{equation} \label{Burbegin}
\frac{\partial u}{\partial t} +  f'(u)\frac{\partial u}{\partial x}  - \nu \frac{\partial^2 u}{\partial x^2} = 0,\quad \nu>0,\ x \in S^1=\R/\Z
\end{equation}
is a popular model for the Navier-Stokes equation, since both of them have similar nonlinearities and dissipative terms (the classical Burgers equation \cite{Bur} corresponds to $f(u)=u^2/2$). For $\nu \ll 1$ and $f$ strongly convex, i.e. satisfying:
\begin{equation} \label{strconvex}
f''(x) \geq \sigma > 0,\quad x \in \R,
\end{equation}
solutions of (\ref{Burbegin}) display turbulent-like behaviour, called "Burgulence" \cite{BF,BK}. In this paper, we are interested in qualitative and quantitative properties of the Burgulence.
\\ \indent
The mean value in space is a conserved quantity for solutions to (\ref{Burbegin}). Indeed, since $u$ is $1$-periodic in space, we have:
\begin{align} \nonumber
\frac{d}{dt}\int_{S^1}{u(t,x) dx}&=-\int_{S^1}{f'(u(t,x))u_x(t,x) dx}+\nu \int_{S^1}{u_{xx}(t,x) dx}=0.
\end{align}
To simplify presentation, we restrict ourselves to solutions with zero mean value in space:
\begin{equation} \label{zero}
\int_{S^1}{u(t,x) dx}=0,\quad \forall t \geq 0.
\end{equation}
\indent
In \cite{Bir}, Biryuk considered (\ref{Burbegin}) with $f$ satisfying (\ref{strconvex}). He studied solutions $u$ for small values of $\nu$ and obtained the following estimates for  norms in $L_2$ of their $m$-th spatial derivatives:
\begin{align} \label{estBiryuk}
&\Vert u(t)\Vert^2_m \leq C \nu^{-(2m-1)}, \frac{1}{T} \int_{0}^{T}{\Vert u(t)\Vert^2_m} \geq c \nu^{-(2m-1)}, m \geq 1, \nu \leq \nu_0.
\end{align}
Note that the exponents for $\nu$ in lower and upper bounds are the same. For fixed $m$, the constants $\nu_0$, $C$, $c$ and $T$ depend on the deterministic initial condition $u_0$. This dependence cannot be removed. Indeed, (\ref{Burbegin}) is dissipative for the $L_2$ norm of $u$, so no non-trivial lower estimate can hold if we take $0$ as the initial condition. Moreover, as $t \rightarrow +\infty$, the solution of the deterministic Burgers equation tends to $0$ uniformly in $u_0$, so we have no hope of getting a non-trivial lower estimate which would hold uniformly in time. In a recent preprint \cite{BorDet}, we formulate the dependence of the estimates (\ref{estBiryuk}) on $u_0$ in a simpler way.
\\ \indent
To get results which are independent of the initial data and hold uniformly for large enough $t$, a natural idea is to introduce a random force and to estimate ensemble-averaged characteristics of solutions. In the article \cite{Bor}, we have considered the case when $0$ in the right-hand side of (\ref{Burbegin}) is replaced by a random spatially smooth force, "kicked" in time. In this article we consider the equation
\begin{equation} \label{whiteBurgersintro}
\frac{\partial u}{\partial t} + f'(u)\frac{\partial u}{\partial x} - \nu \frac{\partial^2 u}{\partial x^2} = \eta^{\omega},
\end{equation}
where $\eta^{\omega}$ is a random force, white in time and smooth in space. This force corresponds to a scaled limit of "kicked" forces with more and more frequent kicks. All forces that we consider have zero mean value in space.
\\ \indent
Study of Sobolev norms of solutions for nonlinear PDEs with small viscosity (with or without random forcing) in order to get estimates for small-scale quantities such as the spectrum is motivated by the problem of turbulence. This research was initiated by Kuksin, who obtained lower and upper estimates of these norms by negative powers of the viscosity for a large class of equations (see \cite{KukGAFA97,KukGAFA99} and references in \cite{KukGAFA99}), and continued by Biryuk \cite{Bir} for the Burgers equation. We use some methods and ideas from those works. Note that for the Burgers equation considered in \cite{Bir,Bor,BorDet} and in the current paper, estimates on Sobolev norms are asymptotically sharp in the sense that viscosity enters lower and upper bounds at the same negative power. Such estimates are not available for the more complicated equations considered in \cite{KukGAFA97,KukGAFA99}.
\\ \indent
In this work, after introducing the notation and setup in Section~\ref{nota}, we formulate the main results in Section~\ref{results}. In Section~\ref{sob}, we begin by estimating from above the moments of $\max \partial u/\partial x$ for solutions $u(t,x)$ of (\ref{whiteBurgersintro}) for $t \geq 1$. Using these bounds, we obtain estimates of the same type as in \cite{Bir,Bor}, valid for time $t \geq T_1=T_0+2$. Here, $T_0$ is a constant, independent of the initial condition and of $\nu$. Actually, for $t \geq T_1$, we are in a  quasi-stationary regime: all estimates hold uniformly in $t,\nu$ and in the initial condition $u_0$.
\\ \indent
In Section~\ref{turb} we study implications of our results in terms of the theory of Burgulence.  Namely, we give sharp upper and lower bounds for the dissipation length scale, increments, flatness and spectral asymptotics for the flow $u(t,x)$ for $t \geq T_1$. These bounds hold uniformly in $\nu \leq \nu_0$, where $\nu_0$ is a positive constant which is independent of $u_0$.
\\ \indent
The results of Section~\ref{turb} rigorously justify the physical predictions for space increments of solutions $u(t,x)$ and for their spectral asymptotics \cite{AFLV,Cho,EKMS2,Kid,Kra}. Our proof of Theorem~\ref{lowerinert} in this section uses an argument from \cite{AFLV}. Note that predictions for spectral asymptotics have been known since the 1950s: in \cite{Kra}, the author refers to some earlier results by Burgers and Tatsumi.
\\ \indent
The rigorous proof of the asymptotics predicted by a physical argument, even for such a relatively simple model as the stochastic Burgers equation, is important since for the 3D or 2D incompressible Navier-Stokes equation there is no exact theory of this type, corresponding to the heuristic theories due to Kolmogorov and Kraichnan. Note that since we study the generalised equation (\ref{whiteBurgersintro}) and not only the equation with the classical nonlinearity $uu_x$, we cannot use the Cole-Hopf transformation \cite{Col,Hop}.
\\ \indent
In Section~\ref{stat}, we prove that the stochastic Burgers equation admits a unique stationary measure $\mu$, and we estimate the speed of convergence to $\mu$ as $t \rightarrow +\infty$. It follows that the estimates in Sections~\ref{sob}-\ref{turb} still hold if we replace averaging in time and probability with averaging with respect to $\mu$.
\\ \indent
We are concerned with solutions for (\ref{whiteBurgersintro}) with small but positive $\nu$. For a detailed study of the limiting dynamics with $\nu=0$, see \cite{EKMS}. Additional properties for the limit corresponding to $t \rightarrow +\infty$ in both cases $\nu=0$ and $\nu>0$ have been established in \cite{GIKP,IK}.
\\ \indent
The results of Sections~\ref{turb}-\ref{stat} also hold in the case of a "kicked" force, for which we have estimates analogous to those in Section~\ref{sob} \cite{Bor}. We would also like to note that similar estimates hold in the case of the multidimensional potential randomly forced Burgers equation (see \cite{BK} for the physical predictions). Those estimates will be the subject of a future publication.

\section{Notation and setup} \label{nota}

\textbf{Agreement}: In the whole paper, all functions that we consider are real-valued.

\subsection{Sobolev spaces}

Consider a zero mean value integrable function $v$ on $S^1$. For $p \in [1,\infty]$, we denote its $L_p$ norm by $\left|v\right|_p$. The $L_2$ norm is denoted by  $\left|v\right|$, and $\left\langle \cdot,\cdot\right\rangle$ stands for the $L_2$ scalar product. From now on $L_p,\ p \in [1,\infty],$ denotes the space of zero mean value functions in $L_p(S^1)$. Similarly, $C^{\infty}$ is the space of $C^{\infty}$-smooth zero mean value functions on $S^1$.
\\ \indent
For a nonnegative integer $m$ and $p \in [1,\infty]$, $W^{m,p}$ stands for the Sobolev space of zero mean value functions $v$ on $S^1$ with finite homogeneous norm
\begin{equation} \nonumber
\left|v\right|_{m,p}=\left|\frac{d^m v}{dx^m}\right|_p.
\end{equation}
In particular, $W^{0,p}=L_p$ for $p \in [1,\infty]$. For $p=2$, we denote $W^{m,2}$ by $H^m$ and abbreviate the corresponding norm as $\left\|v\right\|_m$. 
\\ \indent
Note that since the length of $S^1$ is $1$, we have
$$
|v|_1 \leq |v|_{\infty} \leq |v|_{1,1} \leq |v|_{1,\infty} \leq \dots \leq |v|_{m,1} \leq |v|_{m,\infty} \leq \dots
$$
We recall a version of the classical Gagliardo-Nirenberg inequality (see \cite[Appendix]{DG}):
\begin{lemm} \label{GN}
For a smooth zero mean value function $v$ on $S^1$,
$$
\left|v\right|_{\beta,r} \leq C \left|v\right|^{\theta}_{m,p} \left|v\right|^{1-\theta}_{q},
$$
where $m>\beta\geq 0$, and $r$ is defined by
$$
\frac{1}{r}=\beta-\theta \Big( m-\frac{1}{p} \Big)+(1-\theta)\frac{1}{q},
$$
under the assumption $\theta=\beta/m$ if $p=1$ or $p=\infty$, and $\beta/m \leq \theta < 1$ otherwise. The constant $C$ depends on $m,p,q,\beta,\theta$.
\end{lemm}
\indent
For any $s \geq 0$, $H^{s}$ stands for the Sobolev space of zero mean value functions $v$ on $S^1$ with finite norm
\begin{equation} \label{Sobolevspectr}
\left\|v\right\|_{s}=(2 \pi)^{s} \Big( \sum_{k \in \Z}{|k|^{2s} |\hat{v}^k|^2} \Big)^{1/2},
\end{equation}
where $\hat{v}^k$ are the complex Fourier coefficients of $v(x)$. For an integer $s=m$, this norm coincides with the previously defined $H^m$ norm. For $s \in (0,1)$, $\left\|v\right\|_{s}$ is equivalent to the norm
\begin{equation} \label{Sobolevfrac}
\left\|v\right\|^{'}_{s}=\Bigg( \int_{S^1} \Big(\int_0^1 {\frac{|v(x+\ell)-v(x)|^2}{\ell^{2s+1}} d \ell} \Big) dx \Bigg)^{1/2}
\end{equation}
(see \cite{Ada,Tay1}).
\\ \indent
Subindices $t$ and $x$, which can be repeated, denote partial differentiation with respect to the corresponding variables. We denote by $v^{(m)}$ the $m$-th derivative of $v$ in the variable $x$. For shortness, the function $v(t,\cdot)$ is denoted by $v(t)$.

\subsection{Random setting} \label{rand}

We provide each space $W^{m,p}$ with the Borel $\sigma$-algebra. Then we consider
an $L_2$-valued Wiener process 
$$
w(t)=w^{\omega}(t),\ \omega \in \Omega,\ t \geq 0,
$$
defined on a complete probability space $(\Omega,\ \F,\ \Pe)$, and the corresponding filtration $\left\{ \F_t,\ t \geq 0 \right\}$. We assume that for each $m$ and each $t \geq 0$, $w(t) \in H^m$, almost surely. That is, for $\zeta,\chi \in L_2,$
$$
\E(\left\langle w(s),\zeta\right\rangle \left\langle w(t),\chi \right\rangle)=\min(s,t) \left\langle Q\zeta,\chi\right\rangle,
$$
where $Q$ is a symmetric operator which defines a continuous mapping $Q: L_2 \rightarrow H^m$ for every $m$. Thus, $w(t) \in C^{\infty}$ for every $t$, almost surely. From now on, we redefine the Wiener process so that this property holds for all $\omega \in \Omega$. We will denote $w(t)(x)$ by $w(t,x)$. For $m \geq 0$, we denote by $I_m$ the quantity
$$
I_m=Tr_{H^m}(Q)=\E \left\|w(1)\right\|_m^2.
$$
For more details on Wiener processes in Hilbert spaces, see \cite[Chapter 4]{DZSto} and \cite{Kuo}.
\\ \indent
For instance, we can consider the "diagonal" Wiener process:
\begin{equation} \nonumber
w(t)=\sqrt{2} \sum_{k \leq -1}{b_k w_k(t) \cos(2 \pi kx)}+\sqrt{2}\sum_{k \geq 1}{b_k w_k(t) \sin(2 \pi kx)},
\end{equation}
where $w_k(t),\ k \neq 0,$ are standard independent Wiener processes and for every $m \geq 0$,
$$
I_m=\sum_{k \geq 1}{b_k^2 (2 \pi k)^{2m}} < \infty.
$$
From now on, the term $dw(s)$ denotes the stochastic differential corresponding to the Wiener process $w(s)$ in the space $L_2$.
\\ \indent
Now fix $m \geq 0$. By Fernique's Theorem \cite[Theorem 3.3.1]{Kuo}, there exist $\lambda_{m},C_m>0$ such that
\begin{equation} \label{eexp}
\E \exp \Big(\lambda_m \left\|w(T)\right\|_m^2/T \Big) \leq C_m,\quad T \geq 0.
\end{equation}
Therefore by Doob's maximal inequality for infinite-dimensional submartingales \cite[Theorem 3.8. (ii)]{DZSto} we have
\begin{equation} \label{moments}
\E \sup_{t \in [0,T]} {\left\|w(t)\right\|^p_m} \leq \Big( \frac{p}{p-1} \Big)^p \E \left\|w(T)\right\|_m^p < +\infty,
\end{equation}
for any $T>0$ and $p \in (1,\infty)$.
Moreover, applying Doob's maximal inequality to $\exp(\alpha \left\|w(T)\right\|_m)$ and maximising in $\alpha$, we prove the existence of $C'_m>0$ such that
\begin{equation} \label{Gauss}
\Pe (\sup_{t \in [0,T]} {\left\|w(t)\right\|_m} \geq \lambda) \leq \exp(-\lambda^2/2C'_mT),\quad T,\lambda>0.
\end{equation}
Note that analogues of (\ref{moments}-\ref{Gauss}) still hold, uniformly in $\tau$, if we replace $\sup_{t \in [0,T]} {\left\|w(t)\right\|_m}$ by $\sup_{t \in [\tau,T+\tau]} {\left\|w(t)-w(\tau)\right\|_m}$ .

\subsection{Preliminaries} \label{prel}

\indent
We begin by considering the free Burgers-type parabolic equation (\ref{Burbegin}). Here, $t \geq 0$, $x \in S^1=\R/\Z$ and the viscosity coefficient satisfies $\nu \in (0,1]$. The function $f$ is $C^{\infty}$-smooth and strongly convex, i.e. it satisfies (\ref{strconvex}). We also assume that its derivatives satisfy:
\begin{equation} \label{poly}
\forall m \geq 0,\ \exists h \geq 0,\ C_m>0:\ |f^{(m)}(x)| \leq C_m (1+|x|)^h,\quad x \in \R,
\end{equation}
where $h=h(m)$ is a function such that $1 \leq h(1) < 2$ (the lower bound on $h(1)$ follows from (\ref{strconvex})). The usual Burgers equation corresponds to $f(x)=x^2/2$.
\\ \indent
The white-forced generalised Burgers equation is (\ref{whiteBurgersintro}) with $\eta^{\omega}=\partial w^{\omega}/\partial t$, where $w^{\omega}(t),\ t \geq 0,$ is the Wiener process defined above. 

\begin{defi} \label{defH1}
We say that an $H^1$-valued process $u(t,x)=u^{\omega}(t,x)$ is a solution of the equation
\begin{equation} \label{whiteBurgers}
\frac{\partial u^{\omega}}{\partial t} + f'(u^{\omega})\frac{\partial u^{\omega}}{\partial x} - \nu \frac{\partial^2 u^{\omega}}{\partial x^2} = \eta^{\omega}
\end{equation}
for $t \geq T$ if:
\\ \indent
(i) For every $t \geq T$, $\omega \mapsto u^{\omega}(t,\cdot)$ is $\F_t$-measurable.
\\ \indent
(ii) For every $\omega$ and for $t \geq T$, $t \mapsto u^{\omega}(t,\cdot)$ is continuous in $H^1$ and satisfies
\begin{align} \nonumber
u^{\omega}(t)=&u^{\omega}(T)-\int_{T}^{t}{\Big( \nu Lu^{\omega}(s)+\frac{1}{2} B(u^{\omega})(s) \Big) ds}
\\ \label{Burgersintbis}
&+w^{\omega}(t)-w^{\omega}(T),
\end{align}
where
$$
B(u)=2 f'(u)u_x;\quad L=-\partial_{xx}.
$$
For shortness, solutions for $t \geq 0$ will be referred to as solutions.
\end{defi}

When studying solutions of (\ref{whiteBurgers}), we always assume that the initial condition $u_T=u(T,\cdot)$ is $\F_T$-measurable and (except in Section~\ref{stat}) that $T=0$ and the initial condition is $C^{\infty}$-smooth. For a given $u_T$, (\ref{whiteBurgers}) has a unique solution, i.e. any two solutions coincide for every $\omega$. For shortness, this solution will be denoted by $u$. This is proved using a straightforward modification of the arguments in \cite[Chapter 14]{DZErg}.
\\ \indent
Since the forcing and the initial condition are smooth in space, the mapping $t \mapsto u(t)$ is time-continuous in $H^m$ for every $m$, and $t \mapsto u(t)-w(t)$ has a space derivative in $C^{\infty}$ for all $t$. In this paper, we always assume that $u_T$ satisfies (\ref{zero}). Consequently, since the mean value of $w(t)$ vanishes identically, $u(t)$ also satisfies (\ref{zero}) for all times.
\\ \indent
Solutions of (\ref{whiteBurgers}) make a time-continuous Markov process in $H^1$. For details, we refer to \cite{KuSh}, where a white force is introduced in a similar setting.
\\ \indent
Now consider, for a solution $u(t,x)$ of (\ref{whiteBurgers}), the functional $G_m(u(t))=\left\|u(t)\right\|_m^2$ and apply It{\^o}'s formula \cite[Theorem 4.17]{DZSto} to (\ref{Burgersintbis}):
\begin{align} \nonumber
\left\|u(t)\right\|_m^2=& \left\|u_T\right\|_m^2 - \int_{T}^{t}{\left( 2\nu \left\|u(s)\right\|_{m+1}^2 +\langle L^m u(s),\ B(u)(s) \rangle\right) ds}
\\ \label{Itoexpint}
 &+ (t-T)I_m + 2 \int_{t}^{T}{\langle L^m u(s),\ dw(s)\rangle}
\end{align}
(we recall that $I_m=Tr(Q_m)$.) Consequently,
\begin{align} \label{Itoexpdiff}
\frac{d}{dt} \E \left\|u(t)\right\|_m^2 &=-2 \nu \E \left\|u(t)\right\|_{m+1}^2 - \E\ \langle L^m u(t),\ B(u)(t) \rangle+ I_m.
\end{align}
As $\langle u,\ B(u) \rangle=0$, for $m=0$ this relation becomes
\begin{align} \label{Itoexp0diff}
\frac{d}{dt} \E \left|u(t)\right|^2 &= I_0-2 \nu \E \left\|u(t)\right\|_{1}^2.
\end{align}

\subsection{Agreements} \label{agree}

From now on, all constants denoted by $C$ with sub- or superindexes are positive and nonrandom. Unless otherwise stated, they depend only on $f$ and on the distribution of the Wiener process $w$. Moreover, all quantities in the paper implicitly depend on those two parameters. By $C(a_1,\dots,a_k)$ we denote constants which also depend on parameters $a_1,\dots,a_k$. By $X \overset{a_1,\dots,a_k}{\lesssim} Y$ we mean that
$$
X \leq C(a_1,\dots,a_k) Y.
$$
The notation $X \overset{a_1,\dots,a_k}{\sim} Y$ stands for
$$
Y \overset{a_1,\dots,a_k}{\lesssim} X \overset{a_1,\dots,a_k}{\lesssim} Y.
$$
In particular, $X \lesssim Y$ and $X \sim Y$ mean that $X \leq C Y$ and $C^{-1} Y \leq X \leq C Y$, respectively. All constants are independent of the viscosity $\nu$ and of the initial value $u_0$.
\\ \indent
We denote by $u=u(t,x)$ a solution of (\ref{whiteBurgers}) with an initial condition $u_0$. For simplicity, in Sections~\ref{sob}-\ref{turb}, we assume that $u_0$ is deterministic. However, we can easily generalise all results to the case of an $\F_0$-measurable random initial condition independent of $w(t), t \geq 0$. Indeed, for any measurable functional $\Phi(u(\cdot))$ we have
$$
\E \Phi(u(\cdot))=\int{\E \Big(\Phi(u(\cdot))|u(0)=u_0 \Big) \mu(du_0)},
$$
where $\mu(u_0)$ is the law of $u_0$, and all estimates in Sections~\ref{sob}-\ref{turb} hold uniformly in $u_0$.
\\ \indent
Moreover, for $\tau \geq 0$ and $u_0$ independent of $w(t)-w(\tau),\ t \geq \tau$, the Markov property yields:
$$
\E \Phi(u(\cdot))=\int{\E \Big(\Phi(u(\tau+\cdot))|u(\tau)=u_0 \Big) \mu(du_0)}.
$$
Consequently, all estimates which hold for time $t$ or a time interval $[t,t+T]$ for solutions $u(t)$ to (\ref{whiteBurgers}) actually hold for time $t+\tau$ or a time interval $[t+\tau,t+\tau+T]$ for $u(t)$ which solves (\ref{whiteBurgers}) for $t \geq \tau$, uniformly in $\tau \geq 0$.
\\ \indent
We use the notation $g^{-}=\max(-g,0)$ and $g^{+}=\max(g,0)$.
\\ \indent
For $T_2 >T_1 \geq 0$ and a Sobolev space $W^{m,p}$, we denote by
\\
$C(T_1,T_2;W^{m,p})$ the space of continuous functions $v$ from $[T_1,T_2]$ to $W^{m,p}$ equipped with the norm $\sup_{s \in [T_1,T_2]}{|v(s)|_{m,p}}$.
%
%

\subsection{Setting and notation in Section~\ref{turb}} \label{agreeturb}

For an observable $A$, i.e. a real-valued functional on a Sobolev space $H^m$, which we evaluate on the solutions $u^{\omega}(s)$, we denote by $\lbrace A \rbrace$ the average of $A(u^{\omega}(s))$ in ensemble and in time over $[t,t+T_0]$:
$$
\lbrace A \rbrace=\frac{1}{T_0}\ \int_{t}^{t+T_0}{\E A(u^{\omega}(s)) ds},\ t \geq T_1=T_0+2.
$$
The constant $T_0$ is the same as in Theorem~\ref{avoir}.
\\ \indent
In this section, we assume that $\nu \leq \nu_0$, where $\nu_0$ is a positive constant. Next, we define the intervals
\begin{equation} \label{ranges}
J_1=(0,\ C_1 \nu];\ J_2=(C_1 \nu,\ C_2];\ J_3=(C_2,\ 1].
\end{equation}
In other words, $J_1 = \lbrace \ell:\ 0 < \ell \lesssim \nu \rbrace$, $J_2 = \lbrace \ell:\ \nu \lesssim \ell \lesssim 1 \rbrace$, $J_3=\lbrace \ell:\ \ell \sim 1\rbrace$. For the values of $\nu_0$, $C_1$ and $C_2$, see (\ref{nu0eq}).
\\ \indent 
In terms of the Kolmogorov 1941 theory \cite{Fri}, the interval $J_1$ corresponds to the \textit{dissipation range}, i.e. for the Fourier modes $k$ such that $|k|^{-1} \preceq C_1 \nu$, $\lbrace|\hat{u}^k|^2\rbrace$ decreases super-algebraically in $k$. The interval $J_2$ corresponds to the \textit{inertial range}, where layer-averaged quantities such as the \textit{energy spectrum} $E(k)$ defined by:
\begin{equation} \label{spectrum}
E(k)=\Bigg\{ \frac{\sum_{|n| \in [M^{-1}k,Mk]}{|\hat{u}^n|^2}}{\sum_{|n| \in [M^{-1}k,Mk]}{1}} \Bigg\}
\end{equation}
behave as a negative degree of $k$. Here $M \geq 1$ is a large enough constant (cf. the proof of Theorem~\ref{spectrinert}). The boundary $C_1 \nu$ between these two ranges is the \textit{dissipation length scale}. Finally, the interval $J_3$ corresponds to the \textit{energy range}, i.e. the sum $\Sigma \lbrace| \hat{u}^k|^2\rbrace$ is mostly supported by the Fourier modes corresponding to $|k|^{-1} \in J_3$.
Actually the positive constants $C_1$ and $C_2$ can take any value, provided
\begin{equation} \label{C1C2}
C_1 \leq \frac{1}{4} K^{-2};\quad 5 K^2 \leq \frac{C_1}{C_2}<\frac{1}{\nu_0}.
\end{equation}
Here, $K$ is a positive constant, chosen in (\ref{K}). Note that the intervals defined by (\ref{ranges}) are non-empty and do not intersect each other for all values of $\nu \in (0,\nu_0]$, under the assumption (\ref{C1C2}). 
\\ \indent
By Theorem~\ref{avoir} we have $\lbrace |u|^2 \rbrace \sim 1$ and (after integration by parts) $\lbrace |\hat{u}^n|^2 \rbrace \leq \lbrace |u|_{1,1}^2 \rbrace/(2 \pi n)^2 \sim 1/n^2$. We recall that we denote by $\hat{u}^n$ the complex Fourier coefficients of $u$. Thus, the ratio
\begin{equation} \nonumber
\frac{\Sigma_{|n|^{-1} \in J_3} |\hat{u}^n|^2}{\Sigma_{n \in \Z} |\hat{u}^n|^2}
\end{equation}
tends to $1$ as $C_2$ tends to $0$, uniformly in $\nu$. Since there exist couples $(C_1,C_2)$ satisfying (\ref{C1C2}) such that $C_2$ is as small as desired, we may for instance assume that
$$
\sum_{|n| < C_2^{-1}}{\lbrace|\hat{u}^n|^2\rbrace} \geq \frac{99}{100} \sum_{n \in \Z}{\lbrace|\hat{u}^n|^2\rbrace}.
$$
For $p,\alpha \geq 0$, we consider the quantity
\begin{equation} \nonumber
S_{p,\alpha}(\ell) = \Big\{ \Big( \int_{S^1}{|u(x+\ell)-u(x)|^p dx} \Big)^{\alpha} \Big\}.
\end{equation}
The quantity $S_{p,1}(\ell)$ is denoted by $S_p(\ell)$: it corresponds to the structure function of $p$-th order, while the flatness $F(\ell)$, given by
\begin{equation} \label{flatness}
F(\ell)=S_4(\ell)/S_2^2(\ell),
\end{equation}
measures spatial intermittency (see \cite{Fri}).

\section{Main results} \label{results}

In Section~\ref{sob}, we prove sharp upper and lower estimates for a large class of Sobolev norms of $u$. A key result is proved in Theorem~\ref{uxpos}. Namely, there we obtain that for $k \geq 1$,
\begin{equation} \label{uxposresults}
\E\ \big( \max_{s \in [t,t+1]} \max_{x \in S^1} u_x(s,x) \big)^{k} \overset{k}{\lesssim} 1,\quad t \geq 1.
\end{equation}
The main estimates are those in the first part of Theorem~\ref{avoir}. There we prove that for $m \in \lbrace 0,1 \rbrace$ and $p \in [1,\infty]$ or for $m \geq 2$ and $p \in (1,\infty]$,
\begin{equation} \label{avoirresults}
\Big( \frac{1}{T} \int_{t}^{t+T}{\E \left|u(s)\right|_{m,p}^{\alpha}} \Big)^{1/\alpha} \overset{m,p,\alpha}{\sim} \nu^{-\gamma},\quad \alpha>0,\ t \geq T_0+2,\ T \geq T_0,
\end{equation}
where $\gamma=\max (0,m-1/p)$, and $T_0$ is a constant.
\\ \indent
In Section~\ref{turb} we assume that $\nu \in (0,\nu_0]$, where $\nu_0 \in (0,1]$ is a constant. Then, we obtain sharp estimates for analogues of quantities characterising hydrodynamical turbulence. Although we only prove results for quantities averaged over a time period of length $T_0$, those results can be immediately extended to quantities averaged over time periods of length $T \geq T_0$. 
\\ \indent
As the first application of estimates (\ref{uxposresults}-\ref{avoirresults}), in Section~\ref{turb} we obtain sharp estimates for the quantities $S_{p,\alpha},\ \alpha \geq 0$. Namely, by Theorem~\ref{avoir2}, for $\ell \in J_1$:
$$
\quad \ \ \ S_{p,\alpha}(\ell) \overset{p,\alpha}{\sim} \left\lbrace \begin{aligned} & \ell^{\alpha p},\ 0 \leq p \leq 1. \\ & \ell^{\alpha p} \nu^{-\alpha(p-1)},\ p \geq 1, \end{aligned} \right.$$
and on the other hand for $\ell \in J_2$:
$$
S_{p,\alpha}(\ell) \overset{p,\alpha}{\sim} \left\lbrace \begin{aligned} & \ell^{\alpha p},\ 0 \leq p \leq 1. \\ & \ell^{\alpha},\ p \geq 1. \end{aligned} \right.
$$
Consequently, for $\ell \in J_2$ the flatness function $F(\ell)=S_4(\ell)/S_2^2(\ell)$ satisfies $F(\ell) \sim \ell^{-1}.$ Thus, solutions $u$ are highly intermittent in the inertial range (see \cite{Fri}).
\\ \indent
On the other hand, we obtain estimates for the spectral asymptotics of Burgulence. Namely, for all $m \geq 1$ and $k \in \Z,\ k \neq 0$ we have:
$$
\lbrace |\hat{u}^k|^2 \rbrace \overset{m}{\lesssim} (k \nu)^{-2m} \nu,
$$
and by Theorem~\ref{spectrinert} and Remark~\ref{spectrinertrmq} for $k$ such that $k^{-1} \in J_2$ we have:
$$
\Bigg\{ \Bigg( \frac{\sum_{|n| \in [M^{-1}k,Mk]}{|\hat{u}^n|^2}}{\sum_{|n| \in [M^{-1}k,Mk]}{1}} \Bigg)^{\alpha} \Bigg\} \overset{\alpha}{\sim} k^{-2\alpha},\quad \alpha>0,
$$
for large enough values of $M>1$. In particular, in the inertial range the energy spectrum satisfies $E(k) \sim k^{-2}$.
\\ \indent
Finally, in Section~\ref{stat}, we prove that (\ref{whiteBurgers}) admits a unique stationary measure $\mu$. Consequently, all upper and lower estimates listed above still hold if we redefine the brackets as averaging with respect to $\mu$, i.e.
$$
\left\{ f(u) \right\}=\int{f(u) \mu(du)}.
$$
Moreover, as $t \rightarrow +\infty$, the rate of convergence to $\mu$ in the Lipschitz-dual distance for Borel probability measures on $L_1$ is at least of the form $C t^{-1/13}$, where $C$ does not depend on the initial condition or on the viscosity $\nu$.

\section{Estimates for Sobolev norms} \label{sob}

\subsection{Upper estimates} \label{sobupper}

The following theorem is proved using a stochastic version of the Kruzhkov maximum principle (cf. \cite{Kru}).

\begin{theo} \label{uxpos}
Denote by $X_t$ the random variable
$$
X_t=\max_{s \in [t,t+1]} \max_{x \in S^1} u_x(s,x).
$$
For every $k \geq 1$, we have
$$
\E \ X_t^{k} \overset{k}{\lesssim} 1,\quad t \geq 1.
$$
\end{theo}

\textbf{Proof.}
We take $t=1$, denoting $X_t$ by $X$: the general case follows by the argument exposed in Subsection~\ref{agree}.
\\ \indent
Consider the equation (\ref{whiteBurgers}) on the time interval $[0,2]$. Putting $v=u-w$ and differentiating once in space,  we get
\begin{equation} \label{maxvx}
\frac{\partial v_x}{\partial t} + f''(u)(v_x+w_x)^2+f'(u) (v_{x}+w_{x})_x=\nu (v_{x}+w_{x})_{xx}.
\end{equation}
Consider $\tilde{v}(t,x)=tv_x(t,x)$ and multiply (\ref{maxvx}) by $t^2$. For $t>0$, $\tilde{v}$ satisfies
\begin{align} \nonumber
&t\tilde{v}_t -\tilde{v} + f''(u) (\tilde{v}+tw_x)^2 + tf'(u) \tilde{v}_x + t^2 f'(u) w_{xx}
\\ \label{maxtildev}
&= \nu t \tilde{v}_{xx}+\nu t^2 w_{xxx}.
\end{align}
Now observe that if the zero mean function $\tilde{v}$ does not vanish identically on the domain $S=\left[0,2\right] \times S^1$, then it attains its positive maximum $N$ on $S$ at a point $(t_1,x_1)$ such that $t_1>0$. At $(t_1,x_1)$ we have $\tilde{v}_t \geq 0$, $\tilde{v}_x=0$, and $\tilde{v}_{xx} \leq 0$.
By (\ref{maxtildev}), at $(t_1,x_1)$ we have the inequality
\begin{equation} \label{maxpoint}
f''(u) (\tilde{v}+tw_x)^2 \leq \tilde{v}- t^2 f'(u) w_{xx}+ \nu t^2 w_{xxx}.
\end{equation}
Denote by $A$ the random variable
$$
A=\max_{t \in [0,2]} |w(t)|_{3,\infty}.
$$
Since for every $t$, $tv(t)$ is the zero space average primitive of $\tilde{v}(t)$ on $S^1$, we get
\begin{align} \nonumber
\max_{t \in [0,2],\ x \in S^1}{|tu|} &\leq \max_{t \in [0,2],\ x \in S^1}{(|tv|+|tw|)} 
\\ \nonumber
&\leq N+2\max_{t \in [0,2]} |w(t)|_{\infty} \leq N+2A.
\end{align}
Now denote by $\delta$ the quantity
\begin{equation} \label{delta}
\delta=2-h(1).
\end{equation}
(cf. (\ref{poly})). Since $\delta>0$, we obtain that
\begin{align} \nonumber
\max_{t \in [0,2],\ x \in S^1} |t^2 f'(u) w_{xx}| & \leq A \max_{t \in [0,2],\ x \in S^1} {t^{\delta} |t^{2-\delta} f'(u)|}
\\ \nonumber
&\leq A \max_{t \in [0,2],\ x \in S^1} {t^{\delta} (|tu|+t)^{2-\delta}}
\\ \nonumber
&\leq C A (N+2A+2)^{2-\delta}.
\end{align}
From now on, we assume that $N \geq 2A$. Since $\nu \in (0,1]$ and $f'' \geq \sigma$, the relation (\ref{maxpoint}) yields
$$
\sigma (N-2A)^2 \leq N+C A (N+2A+2)^{2-\delta}+4A.
$$
Thus we have proved that if $N \geq 2A$, then $N \leq C(A+1)^{1/\delta}$. Since by (\ref{moments}), all moments of $A$ are finite, all moments of $N$ are also finite. By definition of $\tilde{v}$ and $S$, the same is true for $X$. This proves the theorem's assertion. $\qed$
\begin{rmq} \label{rmq11}
Actually, using (\ref{Gauss}), we can prove that there exist $\beta,\beta'>0$ such that
$$
\E \exp(\beta X_t^{2 \delta}) \leq \E \exp \Big(\beta' (\max_{t \in [0,2]} |w(t)|_{3,\infty}+1)^2 \Big) \lesssim 1,\quad t \geq 1.
$$
\end{rmq}

\begin{cor} \label{W11}
For $k \geq 1$,
$$
\E \max_{s \in [t,t+1]} \left|u(s)\right|^k_{1,1} \overset{k}{\lesssim} 1,\quad t \geq 1.
$$
\end{cor}

\textbf{Proof.}
The space average of $u_x(s)$ vanishes identically. Therefore
$$
\int_{S^1}{\left| u_x(s)\right|}=2 \int_{S^1}{(u_x(s))^{+}} \leq 2 \max_{x \in S^1} u_x(s,x).\ \qed
$$

\begin{cor} \label{Lpupper}
For $k \geq 1$,
$$
\E \max_{s \in [t,t+1]} \left|u(s)\right|^k_{p} \overset{k}{\lesssim} 1,\quad p \in [1,\infty],\ t \geq 1.
$$
\end{cor}

Now we recall a standard estimate of the nonlinearity $\left\langle L^m u, B(u)\right\rangle$ (see Subsection~\ref{prel} for the definitions of $L$ and $B$).

\begin{lemm} \label{lmubuinfty}
For every $m \geq 1$ there exist $C_m>0$ and a natural number $n'=n'(m)$ such that for $w \in C^{\infty}$,
\begin{align} \label{polyestimate}
&N_m(w)=\left| \left\langle L^m w, B(w) \right\rangle\right| \leq C_m (1+\left|w\right|_{\infty} )^{n'} \left\|w\right\|_m \left\|w\right\|_{m+1}.
\end{align}
\end{lemm}

\textbf{Proof.}
Fix $m \geq 1$. Denote $\left|w\right|_{\infty}$ by $N$. Let $C'$ denote various expressions of the form $C_m (1+N)^{n(m)}$. We have
\begin{align} \nonumber
N_m(w) &=  2 \left|\left\langle w^{(2m)}, (f(w))^{(1)} \right\rangle\right|= 2 \left|\left\langle w^{(m+1)}, (f(w))^{(m)} \right\rangle\right|
\\ \nonumber
\leq & C(m) \sum_{k=1}^m\ \sum_{\substack{1 \leq a_1 \leq \dots \leq a_k \leq m \\ a_1+ \dots+a_k = m}} \int_{S^1}{\left| w^{(m+1)} w^{(a_1)} \dots w^{(a_k)} f^{(k)}(w) \right|}
\\ \nonumber
\leq & C(m) \max_{x \in [-N,N]}\ \max(f'(x),\dots f^{(m)}(x))
\\ \nonumber
& \times \sum_{k=1}^m\ \sum_{\substack{1 \leq a_1 \leq \dots \leq a_k \leq m \\ a_1+ \dots+a_k = m}} \int_{S^1} | w^{(a_1)} \dots w^{(a_k)} w^{(m+1)} |.
\end{align}
Using first (\ref{poly}), then H{\"o}lder's inequality, and finally Lemma~\ref{GN}, we get
\begin{align} \nonumber
N_m(w) \leq & C(m) (1+N)^{\max(h(1),\dots,h(m))} 
\\ \nonumber
&\times \sum_{k=1}^m\ \sum_{\substack{1 \leq a_1 \leq \dots \leq a_k \leq m \\ a_1+ \dots+a_k = m}} \int_{S^1} | w^{(a_1)} \dots w^{(a_k)} w^{(m+1)} |
\\ \nonumber
\leq & C' \sum_{k=1}^m\ \sum_{\substack{1 \leq a_1 \leq \dots \leq a_k \leq m \\ a_1+ \dots+a_k = m}} \Big( \left|w^{(a_1)}\right|_{2m/a_1} \dots \left|w^{(a_k)}\right|_{2m/a_k} \left\|w\right\|_{m+1} \Big)
\\ \nonumber
\leq & C' \left\|w\right\|_{m+1} \sum_{k=1}^m\ \sum_{\substack{1 \leq a_1 \leq \dots \leq a_k \leq m \\ a_1+ \dots+a_k = m}} \Big( (\left\|w\right\|_m^{a_1/m} |w|_{\infty}^{(m-a_1)/m} ) \times \dots
\\ \nonumber
& \dots \times (\left\|w\right\|_m^{a_k/m} |w|_{\infty}^{(m-a_k)/m}) \Big)
\\ \nonumber
\leq & C' (1+N)^{m-1} \left\|w\right\|_m \left\|w\right\|_{m+1} = C' \left\|w\right\|_m \left\|w\right\|_{m+1}.\ \qed
\end{align}

\begin{lemm} \label{uppermaux}
For $m \geq 1$,
$$
\E \left\|u(t)\right\|^{2}_m \overset{m}{\lesssim} \nu^{-(2m-1)},\quad t \geq 2.
$$
\end{lemm}

\textbf{Proof.} Fix $m \geq 1$. We will use the notation
\begin{equation} \nonumber
x(s)=\E \left\|u(s)\right\|^2_m;\quad y(s)=\E \left\|u(s)\right\|^2_{m+1}.
\end{equation}
As previously, it suffices to consider the case $t=2$. We claim that for $s \in [1,2]$ we have the implication
\begin{align} \nonumber
x(s) &\geq C' \nu^{-(2m-1)} \Longrightarrow
\\ \label{decrm}
\frac{d}{ds} x(s) &\leq -(2m-1) (x(s))^{2m/(2m-1)},
\end{align}
where $C' \geq 1$ is a fixed number, chosen later. Below, all constants denoted by $C$ are positive and do not depend on $C'$, and we denote by $Z$ the quantity
$$
Z=C' \nu^{-(2m-1)}.
$$
Indeed, assume that $x(s) \geq Z.$ By (\ref{Itoexpdiff}) and Lemma~\ref{lmubuinfty}, we have
\begin{align} \nonumber
\frac{d}{ds} x(s) \leq & - 2 \nu y(s) + C \E \Big( (1+\left|u(s)\right|_{\infty})^{n'} \left\|u(s)\right\|_m \left\|u(s)\right\|_{m+1} \Big)+I_m,
\end{align}
with $n'=n'(m)$. Since by Lemma~\ref{GN} applied to $u_x$, we get
\begin{equation} \label{GN11m}
\left\|u(s)\right\|_m \leq C \left\|u(s)\right\|_{m+1}^{(2m-1)/(2m+1)} \left|u(s)\right|_{1,1}^{2/(2m+1)},
\end{equation}
we obtain that
\begin{align} \nonumber
\frac{d}{ds}  x(s)  \leq & - 2 \nu y(s) + C\E \Big((1+\left|u(s)\right|_{1,1})^{n'+2/(2m+1)}
\\ \nonumber
& \times \left\|u(s)\right\|_{m+1}^{4m/(2m+1)} \Big)+I_m.
\end{align}
Thus by H{\"o}lder's inequality and Corollary~\ref{W11} we get
\begin{align} \nonumber
\frac{d}{ds} x(s) \leq & \Big( - 2 \nu (y(s))^{1/(2m+1)} + C \Big) (y(s))^{2m/(2m+1)}+I_m.
\end{align}
On the other hand, (\ref{GN11m}), H{\"o}lder's inequality and Corollary~\ref{W11} yield
\begin{align} \nonumber
x(s) \leq\ & C ( y(s) )^{(2m-1)/(2m+1)} ( \E |u(s)|_{1,1}^{2} )^{2/(2m+1)}
\\ \nonumber
\leq & C (y(s))^{(2m-1)/(2m+1)},
\end{align}
and thus
\begin{align} \nonumber
( y(s) )^{1/(2m+1)}& \geq C ( x(s) )^{1/(2m-1)}.
\end{align}
Consequently, since $x(s) \geq C' \nu^{-(2m-1)}$, for $C'$ large enough we have
\begin{align} \nonumber
\frac{d}{ds} x(s) \leq &\left(-C C'^{1/(2m-1)}+C \right) ( x(s) )^{2m/(2m-1)}+I_m.
\end{align}
Thus we can choose $C'$ in such a way that (\ref{decrm}) holds.
\\ \indent
Now we claim that
\begin{equation} \label{decrmcor}
x(2) \leq Z.
\end{equation}
Indeed, if $x(s) \leq Z$ for some $s \in \left[1,2\right]$, then the assertion (\ref{decrm}) ensures that $x(s)$ remains below this threshold up to $s=2$: thus we have proved (\ref{decrmcor}).
\\ \indent
Now, assume that $x(s) > Z$ for all $s \in \left[1,2\right]$. Denote
$$
\tilde{x}(s)=(x(s))^{-1/(2m-1)},\ s \in \left[1,2\right].
$$
Using the implication (\ref{decrm}) we get $d\tilde{x}(s)/ds \geq 1$. Therefore $\tilde{x}(2) \geq 1$. As $\nu \leq 1$ and $C' \geq 1$, we get $x(2) \leq Z$. Thus in both cases the inequality (\ref{decrmcor}) holds. This proves the lemma's assertion. $\qed$

\begin{cor} \label{uppermauxcor}
For $m \geq 1$,
$$
\E \left\|u(t)\right\|^{k}_m \overset{m,k}{\lesssim} \nu^{-k(2m-1)/2},\quad k \geq 1,\ t \geq 2.
$$
\end{cor}

\textbf{Proof.}
The cases $k=1,2$ follow immediately from Lemma~\ref{uppermaux}.
\\ \indent
For $k \geq 3$, we consider only the case when $k$ is odd, since the general case follows by H{\"o}lder's inequality. Setting $N=((2m-1)k+1)/2$ and applying Lemma~\ref{GN}, we get
$$
\left\|u(t)\right\|^{k}_m \overset{m,k}{\lesssim} \left\|u(t)\right\|_N \left|u(t)\right|^{k-1}_{1,1}.
$$
Therefore, by H{\"o}lder's inequality, Lemma~\ref{uppermaux} and Corollary~\ref{W11} we get
\begin{align} \nonumber
\E \left\|u(t)\right\|^{k}_m &\overset{m,k}{\lesssim} (\E \left\|u(t)\right\|^2_N)^{1/2} (\E \left|u(t)\right|^{2k-2}_{1,1})^{1/2}
\\ \nonumber
& \overset{m,k}{\lesssim} \nu^{-(N-1/2)} = \nu^{-k(2m-1)/2}.\ \qed
\end{align}

\begin{lemm} \label{uppermlemm}
For $m \geq 1$,
$$
\E  \max_{s \in [t,t+1]} \left\|u(s)\right\|^{2}_m \overset{m}{\lesssim} \nu^{-(2m-1)},\quad t \geq 2.
$$
\end{lemm}

\textbf{Proof.} We begin by fixing $m \geq 1$. As previously, we can take $t=2$. In this proof, the random variables $\Theta_i,\ i \in [1,5]$  are positive and have finite moments. All constants denoted by $C$ are positive and only depend on $m$. We denote $w(t)-w(2)$ by $\tilde{w}(t)$, and $u(t)-\tilde{w}(t)$ by $\tilde{u}(t)$. By (\ref{moments}), it follows that it suffices to prove the result with $u$ replaced by $\tilde{u}$.
\\ \indent
By (\ref{Burgersintbis}), for $s \geq 2$ we have
\begin{align} \nonumber
&\left\|\tilde{u}(s)\right\|^2_m= \left\|\tilde{u}(2)\right\|_m^2 - \int_{2}^{s}{\langle L^m \tilde{u}(s'),\ 2 \nu Lu(s')+B(u(s')) \rangle ds'}
\\ \nonumber
& = \left\|\tilde{u}(2)\right\|_m^2 - \int_{2}^{s}{\langle L^m u(s'),\ 2 \nu Lu(s')+B(u(s')) \rangle ds'}
\\ \nonumber
&+ \int_{2}^{s}{\langle L^m \tilde{w}(s'),\ 2 \nu Lu(s')+B(u(s')) \rangle ds'}
\\ \label{Itolarge}
& = \left\|\tilde{u}(2)\right\|_m^2 - \int_{2}^{s}{\Big( 2 \nu \left\| u(s') \right\|^2_{m+1} + \langle L^m u(s'),\ B(u(s')) \rangle \Big)ds'}
\\ \label{Itosmall}
&+ \int_{2}^{s}{ \Big( 2 \nu \langle L^{m+1} \tilde{w}(s'),\ u(s') \rangle - 2 \langle L^m \tilde{w'}(s'),\ f(u(s')) \rangle \Big) ds'}.
\end{align}
Let
\begin{align} \nonumber
& \Theta_1=1+\max_{s' \in [2,3]} \left|u(s')\right|_{1,1}+\max_{s' \in [2,3],\ x \in S^1} |f(u(s',x))|;
\\ \label{Theta12}
& \Theta_2=1+\max_{s' \in [2,3]} \left| \tilde{w}(s')\right|_{2m+2,1};\ \Theta_3=\nu^{(2m-1)} \left\|\tilde{u}(2)\right\|_m^2.
\end{align}
Using Corollary~\ref{W11}, Corollary \ref{Lpupper} and (\ref{poly}), we obtain that the random variable $\Theta_1$ has all moments finite. Finiteness of moments for $\Theta_2$ follows from (\ref{moments}). Finally, finiteness of moments for $\Theta_3$ follows from Lemma~\ref{uppermaux}, since we have $u(2)=\tilde{u}(2)$. Now denote by $A_1(s)$ and $A_2(s)$ the right-hand sides of (\ref{Itolarge}) and (\ref{Itosmall}), respectively. As in the proof of Lemma~\ref{uppermaux}, by Lemma \ref{lmubuinfty} and Lemma \ref{GN} we get that for $s \in [2,3]$, we have respectively
\begin{align} \nonumber
&| \langle L^m u(s),\ B(u(s))\rangle | \leq C (1+|u(s)|_{\infty})^{n'(m)} \left\|u(s)\right\|_m \left\|u(s)\right\|_{m+1}
\\ \nonumber
&\leq C (1+|u(s)|_{1,1})^{n'(m)} |u(s)|_{1,1}^{2/(2m+1)} \left\|u(s)\right\|^{4m/(2m+1)}_{m+1}
\\ \label{Theta4aux1}
&\leq C \Theta_1^{n'(m)+2/(2m+1)} \left\|u(s)\right\|^{4m/(2m+1)}_{m+1},
\end{align}
and
\begin{align} \label{Theta4aux2}
\left\|u(s)\right\|^2_{m+1} \geq C |u(s)|_{1,1}^{-4/(2m-1)} \left\|u(s)\right\|^{(4m+2)/(2m-1)}_{m}.
\end{align}
Now we claim that there exists a positive random variable of the form
$$
\Theta_4=C \Theta_1^{a(m)}
$$
such that
\begin{align} \label{arrow}
&\left\|u(s)\right\|^2_m \geq \Theta_4 \nu^{-(2m-1)} \Longrightarrow \frac{d A_1(s)}{ds}  \leq 0.
\end{align}
Indeed, by (\ref{Theta4aux2}), if $\left\|u(s)\right\|^2_m \geq \Theta_4 \nu^{-(2m-1)}$, then we have
$$
\left\|u(s)\right\|^2_{m+1} \geq C \Theta_1^{-4/(2m-1)} \Theta_4^{(2m+1)/(2m-1)} \nu^{-(2m+1)},
$$
and therefore by (\ref{Theta4aux1}) we get:
\begin{align} \nonumber
& \frac{d A_1(s)}{ds} = -2\nu \left\|u(s)\right\|_{m+1}^2 - \langle L^m u(s),\ B(u(s))\rangle 
\\ \nonumber
& \leq \left\|u(s)\right\|^{4m/(2m+1)}_{m+1}
(-2\nu \left\|u(s)\right\|^{2/(2m+1)}_{m+1}+C \Theta_1^{n'(m)+2/(2m+1)})  \nu^{-2m}
\\ \nonumber
& \leq \left\|u(s)\right\|^{4m/(2m+1)}_{m+1} (-C \Theta_1^{-4/(4m^2-1)} \Theta_4^{1/(2m-1)}+C \Theta_1^{n'(m)+2/(2m+1)})  \nu^{-2m}.
\end{align}
Moreover, if we define the random variable $\Theta_5$ by
\begin{equation} \nonumber
\Theta_5=(\sqrt{\Theta_4}+\Theta_2)^2,
\end{equation}
then we have
\begin{equation} \label{arrow2}
\left\|\tilde{u}(s)\right\|^2_m \geq \Theta_5 \nu^{-(2m-1)} \Longrightarrow \left\|u(s)\right\|^2_m \geq \Theta_4 \nu^{-(2m-1)}.
\end{equation}
Indeed:
\begin{align} \nonumber
\left\|u(s)\right\|^2_m & = \left\|\tilde{u}(s)+\tilde{w}(s)\right\|^2_m
\\ \nonumber
& \geq \Big(\sqrt{\left\|\tilde{u}(s)\right\|_m^2}-\left\|\tilde{w}(s)\right\|_m \Big)^2.
\end{align}
Now consider the stopping time $\tau$ defined by
$$
\tau=\lbrace \inf s \in [2,3]: \left\|\tilde{u}(s)\right\|^{2}_m \geq \Theta_5 \nu^{-(2m-1)} \rbrace.
$$
By convention, $\tau=3$ if the set in question is empty. Relations (\ref{arrow}-\ref{arrow2}) yield that
\begin{align} \nonumber
&\max_{s \in [2,3]} \left\|\tilde{u}(s)\right\|^{2}_m \leq \left\|\tilde{u}(\tau)\right\|^{2}_m+\max_{s \in [\tau,3]}{A_2(s)}
\\ \label{Itoineq}
&\leq \max(\Theta_3,\Theta_5) \nu^{-(2m-1)} + \int_{s' \in [2,3]} {\Big| \frac{d A_2(s')}{ds'} \Big| ds'}.
\end{align}
To prove the lemma's assertion, it remains to observe that we have:
\begin{align} \nonumber
\int_2^3 {\Big| \frac{d A_2(s')}{ds'} \Big| ds'} \leq& \int_{2}^{3}{ \left(2 \nu \left| \tilde{w}(s')\right|_{2m+2,1} |u(s')|_{\infty}  \right.}
\\ \nonumber
& {\left.  + 2 \left| \tilde{w}(s')\right|_{2m+1,1} \max_{ x \in S^1} |f(u(s',x))|  \right) ds'}
\\ \nonumber
\leq & C \Theta_1 \Theta_2.\ \qed
\end{align}
\medskip
\\ \indent
Repeating the proof of Corollary~\ref{uppermauxcor} we get that for $m \geq 1$,
\begin{equation} \label{upperm}
\E  \max_{s \in [t,t+1]} \left\|u(s)\right\|^{k}_m \overset{m,k}{\lesssim}  \nu^{-k(2m-1)/2},\quad k \geq 1,\ t \geq 2.
\end{equation}
\smallskip
\\ \indent
Denote $\gamma=\max (0,m-1/p)$.

\begin{theo} \label{upperwmp}
For $m \in \lbrace 0,1 \rbrace$ and $p \in [1,\infty]$, or for $m \geq 2$ and $p \in (1,\infty]$,
$$
\Big( \E  \max_{s \in [t,t+1]} \left|u(s)\right|^{\alpha}_{m,p} \Big)^{1/\alpha} \overset{m,p,\alpha}{\lesssim} \nu^{-\gamma},\quad \alpha>0,\ t \geq 2.
$$
\end{theo}

\textbf{Proof.}
We consider only the case when $\alpha$ is an integer: the general case follows by  H{\"o}lder's inequality.
\\ \indent
For $m \geq 1$ and $p \in [2,\infty]$, we interpolate $\left|u(s)\right|_{m,p}$ between $\left\|u(s)\right\|_{m}$ and $\left\|u(s)\right\|_{m+1}$. By Lemma~\ref{GN} we have
$$
\left|u(s)\right|^{\alpha}_{m,p} \overset{p}{\lesssim} (\left\|u(s)\right\|_{m}^{\alpha})^{1-\theta} (\left\|u(s)\right\|_{m+1}^{\alpha})^{\theta},\ \theta=\frac{1}{2}-\frac{1}{p}.
$$
Then we use (\ref{upperm}) and H{\"o}lder's inequality to complete the proof.
\\ \indent
We use the same method to prove the case $m=1,\ p \in [1,2]$, combining the inequality (\ref{upperm}) and Corollary~\ref{W11}. We also proceed similarly for $m \geq 2,\ p \in (1,2)$, combining Corollary~\ref{W11} and an estimate for $\Vert u \Vert_{M,p}^{\alpha}$ for a large value of $M$ and some $p \geq 2$.
\\ \indent
Finally, the case $m=0$ follows from Corollary~\ref{Lpupper}. $\qed$
\medskip
\\ \indent
Unfortunately, the proof of Theorem~\ref{upperwmp} cannot be adapted to the case $m \geq 2$ and $p=1$. Indeed, Lemma~\ref{GN} only allows us to estimate a $W^{m,1}$ norm from above by other $W^{m,1}$ norms: we can only get that
$$
|w|_{m,1} \overset{m,n,k}{\lesssim} |w|_{n,1}^{(m-k)/(n-k)} |w|_{k,1}^{(n-m)/(n-k)},\ 0 \leq k < m < n,
$$
and thus the upper estimates obtained above cannot be used. However, $|u|_{m,1} \leq |u|_{m,1+\beta}$ for any $\beta>0$. Consequently, the theorem's statement holds for $m \geq 2$ and $p=1$, with $\gamma$ replaced by $\gamma+\lambda$, and $\overset{m,p,\alpha}{\lesssim}$ replaced by $\overset{m,p,\alpha,\lambda}{\lesssim}$, for any $\lambda>0$.

\subsection{Lower estimates}

\indent
For a solution $u(t)$ of (\ref{whiteBurgers}), the first quantity that we estimate from below is the expected value of $\frac{1}{T} \int_{t}^{t+T}{\left\|u(s)\right\|_1^2}$, where $t \geq 1$ and $T>0$ is sufficiently large.

\begin{lemm} \label{finitetime}
There exists a constant $T_0>0$ such that we have
$$
\Big( \frac{1}{T} \int_{t}^{t+T}{\ \E \left\|u(s)\right\|_1^2} \Big)^{1/2} \gtrsim \nu^{-1/2},\qquad t \geq 1,\ T \geq T_0.
$$
\end{lemm}

\textbf{Proof.} For $T>0$, by (\ref{Itoexp0diff}) we get
\begin{align} \nonumber
\E \left|u(t+T)\right|^2 &\geq \E ( \left|u(t+T)\right|^2-\left|u(t)\right|^2 )
= TI_0 - 2 \nu \int_{t}^{t+T}{\E \left\|u(s)\right\|_{1}^2}.
\end{align}
On the other hand, by Corollary~\ref{Lpupper} there exists a constant $C'>0$ such that
$\E \left|u(t+T)\right|^2 \leq C'$. Consequently, for $T \geq T_0:=(C'+1)/I_0$,
$$
\frac{1}{T} \int_{t}^{t+T}{\E \left\|u(s)\right\|_1^2} \geq \frac{TI_0-C'}{2T} \nu^{-1} \geq \frac{I_0}{2(C'+1)} \nu^{-1},
$$
which proves the lemma's assertion.\ $\qed$
\medskip \\ \indent
This time-averaged lower bound of the $H^1$ norm yields similar bounds of $H^m$ norms for $m \geq 2$.

\begin{lemm} \label{finalexp}
For $m \geq 1$,
$$
\Big( \frac{1}{T} \int_{t}^{t+T}{\E \left\|u(s)\right\|_m^2} \Big)^{1/2} \overset{m}{\gtrsim} \nu^{-(m-1/2)},\qquad t \geq 1,\ T \geq T_0.
$$
\end{lemm} 

\textbf{Proof.}
Since the case $m=1$ has been treated in the previous lemma, we may assume that $m \geq 2$. By Lemma~\ref{GN}, we have
$$
\Vert u \Vert_1^2 \lesssim \Vert u \Vert_m^{2/(2m-1)} \left|u(s)\right|_{1,1}^2)^{(4m-4)/(2m-1)}.
$$
Therefore by H{\"o}lder's inequality and Corollary~\ref{W11} we get
\begin{align} \nonumber
( \E \left\|u(s)\right\|^2_1 )^{2m-1} &\overset{m}{\lesssim} (\E \left\|u(s)\right\|_m^2) (\E \left|u(s)\right|_{1,1}^2)^{2m-2}
\\ \label{Sobolev11}
&\overset{m}{\lesssim} \E \left\|u(s)\right\|_m^2.
\end{align}
Integrating (\ref{Sobolev11}) in time, we get
\begin{align} \nonumber
\frac{1}{T} \int_{t}^{t+T}{\E \left\|u(s)\right\|_m^2 } \overset{m}{\gtrsim} & \ \frac{1}{T} \int_{t}^{t+T}{(\E \left\|u(s)\right\|^2_1)^{2m-1}}
\\ \nonumber
\overset{m}{\gtrsim} & \ \Big(\frac{1}{T} \int_{t}^{t+T}{\E \left\|u(s)\right\|_1^{2} } \Big)^{2m-1}.
\end{align}
Now the lemma's assertion follows from Lemma~\ref{finitetime}. $\qed$
\smallskip
\\ \indent
The following two results generalise Lemma~\ref{finalexp}. We recall that $\gamma=\max(0,m-1/p)$.

\begin{lemm} \label{finalexpbis}
For $m=0$ and $p=\infty$, or for $m \geq 1$ and $p \in [1,\infty]$,
$$
\Big( \frac{1}{T} \int_{t}^{t+T}{\E \left|u(s)\right|_{m,p}^2} \Big)^{1/2} \overset{m,p}{\gtrsim} \nu^{-\gamma},\quad t \geq 2,\ T \geq T_0.
$$
\end{lemm}

\textbf{Proof.}
In the case $m=1,\ p \geq 2$, it suffices to apply H{\"o}lder's inequality in place of Lemma~\ref{GN} in the proof of an analogue for \\
Lemma~\ref{finalexp}.
\\ \indent
In the case $m \geq 2$, the proof is exactly the same as for Lemma~\ref{finalexp} for $p \in (1,\infty)$. In the cases $p=1,\infty$,  Lemma~\ref{GN} does not allow us to estimate $|u(s)|_{m,p}^2$ from below using $|u(s)|^2_{1,1}$ and $\Vert u(s) \Vert_1^2$. However, for $p=\infty$ we can proceed similarly, using $|u(s)|^2_{\infty}$ and $| u(s) |_{1,\infty}^2$, since for these quantities we already have estimates from above (Corollary~\ref{Lpupper}) and from below, respectively. On the other hand, for $p=1$ it suffices to observe that $\left|u(s)\right|_{m,1} \geq \left|u(s)\right|_{m-1,\infty}$.
\\ \indent
Now consider the case $m=1,\ p \in [1,2)$. By H{\"o}lder's inequality we have
\begin{align} \nonumber
\frac{1}{T} \int_{t}^{t+T}{\E \left|u(s)\right|_{1,p}^2}\geq & \Big( \frac{1}{T} \int_{t}^{t+T}{\E \left\|u(s)\right\|_{1}^2} \Big)^{2/p}
\\ \nonumber
& \times \Big( \frac{1}{T} \int_{t}^{t+T}{\E \left|u(s)\right|_{1,\infty}^2} \Big)^{(p-2)/p}.
\end{align}
Using Lemma~\ref{finitetime} and Theorem~\ref{upperwmp}, we get the lemma's assertion.
\\ \indent
We proceed similarly for the case $m=0,\ p=\infty$. Indeed, by Lemma~\ref{GN} we have
$\left|u(s)\right|_{1,\infty} \leq C \left|u(s)\right|^{1/2}_{\infty} \left|u(s)\right|^{1/2}_{2,\infty}$. Thus, the lemma's assertion follows from H{\"o}lder's inequality, the case $m=1,\ p=\infty$ and  Theorem~\ref{upperwmp} (case $m=2,\ p=\infty$). $\qed$

\begin{lemm} \label{finalexpter}
For $m=0$ and $p=\infty$, or for $m \geq 1$ and $p \in [1,\infty]$,
$$
\Big( \frac{1}{T} \int_{t}^{t+T}{\E \left|u(s)\right|_{m,p}^{\alpha}} \Big)^{1/\alpha} \overset{m,p,\alpha}{\gtrsim} \nu^{-\gamma},\quad \alpha>0,\ t \geq 2,\ T \geq T_0.
$$
\end{lemm}

\textbf{Proof.}
As previously, we may assume that $p>1$. The case $\alpha \geq 2$ follows immediately from Lemma~\ref{finalexpbis} and H{\"o}lder's inequality. The case $\alpha<2$ follows from H{\"o}lder's inequality, the case $\alpha=2$ and Theorem~\ref{upperwmp} (case $\alpha=3$), since we have
\begin{align} \nonumber
\frac{1}{T} \int_{t}^{t+T}{\E \left|u(s)\right|_{m,p}^{\alpha}} \geq &\Big( \frac{1}{T} \int_{t}^{t+T}{\E \left|u(s)\right|_{m,p}^{2}} \Big)^{3-\alpha}
\\ \nonumber
& \times  \Big( \frac{1}{T} \int_{t}^{t+T}{\E \left|u(s)\right|_{m,p}^{3}} \Big)^{\alpha-2}.\ \qed
\end{align}

Now we prove that for every $p \in [1,\infty)$, in a certain sense, $\E |u|_p$ is large if and only if $\E |u|_{\infty}$ is large.

\begin{lemm} \label{Lpequiv}
For $t \geq 1$, denote by $A$ the quantity $\E |u(t)|_{\infty}^2$. Then there exists a constant $C'>0$ such that for $p \in [1,\infty]$ we have
$$
\tilde{g}(A):= \min \Big(\frac{3A}{8}, \frac{3A^2}{16C'} \Big)\leq \E |u(t)|^2_{p} \leq A.
$$
\end{lemm}

\textbf{Proof.} We may take $p=1$. Denote by $l$ the quantity
$$
l=\min (\sqrt{A/2C'},\ 1),
$$
where $C'$ is the upper bound for $\E \ X_t^2$ in the statement of Theorem~\ref{uxpos}. Consider the random point $x=x_t$ where $|u(t,\cdot)|$ reaches its maximum. If this point is not unique, let $x$ be the leftmost such point on $S^1$ considered as $[0,1)$. Let $I$ be the interval $[x,x+l]$ if $u(t,x)<0$, and the interval $[x-l,x]$ if $u(t,x) \geq 0$, respectively. We have
\begin{align} \nonumber
\E |u(t)|^2_{1} & \geq \E\ \Bigg(\int_{I}{|u(t,y)| dy} \Bigg)^2
\\ \nonumber
& \geq \E\ \Bigg( l \Bigg( |u(t)|_{\infty}-\frac{l \max_{x \in S^1} u_x(t)}{2} \Bigg) \Bigg)^2
\\ \nonumber
& \geq l^2 \Bigg( \frac{3}{4} \E |u(t)|^2_{\infty}-\frac{3l^2}{4} \E \Big((\max_{x \in S^1} u_x(t))^2\Big) \Bigg).
\end{align}
By definition of $A$, $C'$ and $l$, we get
\begin{align} \nonumber
\E |u(t)|^2_{1} \geq l^2 \Bigg(\frac{3A}{4}-\frac{3l^2C'}{4} \Bigg) \geq \frac{3l^2A}{8}=\tilde{g}(A). \qed
\end{align}
\smallskip \\ \indent

Finally we prove the following uniform lower estimate.

\begin{lemm} \label{Lplower}
We have
$$
\E |u(t)|_p^2 \gtrsim 1,\quad t \geq T_0+2,\ p \in [1,\infty].
$$
\end{lemm}

\textbf{Proof.}
We can take $p=2$. Indeed, the case $p \in (2,\infty]$ follows immediately from the case $p=2$. On the other hand, the case 
\\
$p \in [1,2)$ follows from H{\"o}lder's inequality, the case $p=2$ and the upper estimate for $\E |u(t)|_{\infty}^2$ in Theorem~\ref{upperwmp}, in the same way as in the proof of Lemma~\ref{finalexpter}.
\\ \indent
Let $C'$ denote various positive constants. From Lemma~\ref{finalexpbis} (case $m=0$ and $p=\infty$), it follows that for some $\tilde{t}$ in $[2,T_0+2]$ we have $\E |u(\tilde{t})|_{\infty}^2 \geq C'$. Then by Lemma~\ref{Lpequiv} we get $\E |u(\tilde{t})|^2 \geq C'$. Thus it suffices to prove that
\begin{equation} \nonumber
\E |u(t)|^2 \leq \kappa \Longrightarrow \frac{d}{dt}{\E |u(t)|^2} \geq 0,\quad t \geq 2,
\end{equation}
where $\kappa$ is a fixed positive number, chosen later.
\\ \indent
If $\E |u(t)|^2 \leq \kappa$, then by Lemma~\ref{Lpequiv}, $\E |u(t)|_{\infty}^2 \leq \tilde{g}^{-1}(\kappa)$. On the other hand, by H{\"o}lder's inequality and Lemma~\ref{GN}, we have
\begin{align} \nonumber
\E \left\|u(t)\right\|_{1}^2 &\leq (\E |u(t)|^2_{1,\infty})^{1/2} (\E |u(t)|^2_{1,1})^{1/2}
\\ \nonumber
& \leq C' (\E |u(t)|^2_{\infty})^{1/4} (\E |u(t)|^2_{2,\infty})^{1/4} (\E |u(t)|^2_{1,1})^{1/2}.
\end{align}
Therefore, by Theorem~\ref{upperwmp}, $\E \left\|u(t)\right\|_{1}^2 \leq C' (\tilde{g}^{-1}(\kappa))^{1/4} \nu^{-1}$, and thus by (\ref{Itoexp0diff}), we get:
$$
\frac{d}{dt}{\E |u(t)|^2} \geq I_0-2C' (\tilde{g}^{-1}(\kappa))^{1/4}.
$$
Since $\tilde{g}^{-1}(\kappa)  \underset{\kappa \to 0}{\longrightarrow} 0$, choosing $\kappa$ small enough so that
$$
2C' (\tilde{g}^{-1}(\kappa))^{1/4} \leq I_0
$$
proves the lemma's assertion. $\qed$
\smallskip \\ \indent
Since $|u(t)|_{1,1} \geq |u(t)|_{\infty}$, an analogue of Lemma~\ref{Lplower} also holds for $|u(t)|_{1,1}$.

\subsection{Main theorem}

\indent
The following theorem sums up the main results of Section~\ref{sob}, with the exception of Theorem~\ref{uxpos}. We recall that $\gamma=\max(0,m-1/p)$.

\begin{theo} \label{avoir}
For $m \in \lbrace 0,1 \rbrace$ and $p \in [1,\infty]$, or for $m \geq 2$ and $p \in (1,\infty]$, we have
\begin{align} \nonumber
&\Big( \frac{1}{T} \int_{t}^{t+T}{\E \left|u(s)\right|_{m,p}^{\alpha}} \Big)^{1/\alpha} \overset{m,p,\alpha}{\sim} \nu^{-\gamma},\ \alpha>0,\ t \geq T_1=T_0+2,
\\ \label{asymp}
& T \geq T_0.
\end{align}
Moreover, the upper estimates hold with time-averaging replaced by maximising over $[t,t+1]$ for $t \geq 2$, i.e.
\begin{equation} \label{maxim}
\Big( \E \max_{s \in [t,t+1]}{\left|u(s)\right|_{m,p}^{\alpha}}\Big)^{1/\alpha} \overset{m,p,\alpha}{\lesssim} \nu^{-\gamma},\quad \alpha>0,\ t \geq 2.
\end{equation}
On the other hand, the lower estimates hold for all $m \geq 0$ and $p \in [1,\infty]$. The asymptotics (\ref{asymp}) hold without time-averaging if $m$ and $p$ are such that $\gamma(m,p)=0$. Namely, in this case,
\begin{equation} \label{gamma0}
\Big( \E \left|u(t)\right|_{m,p}^{\alpha}\Big)^{1/\alpha} \overset{m,p,\alpha}{\sim} 1,\quad \alpha>0,\ t \geq T_1.
\end{equation}
\end{theo}

\textbf{Proof.}
The upper estimates for all cases, as well as the lower estimates in (\ref{asymp}) for all cases and in (\ref{gamma0}) for the case $\alpha=2$, follow from the lemmas and theorems above. For $\alpha>2$, the lower estimates in (\ref{gamma0}) follow immediately from the lower estimates for $\alpha=2$. For $\alpha < 2$, these estimates are obtained from H{\"o}lder's inequality, the lower estimates for $\alpha=2$ and the upper estimates for $\alpha=3$ in the same way as in the proof of Lemma~\ref{finalexpter}.
\medskip
\\ \indent
This theorem yields, for integers $m \geq 1$, the relation
\begin{equation} \label{integers}
\lbrace \Vert u\Vert_m^2 \rbrace \overset{m}{\sim} \nu^{-(2m-1)}.
\end{equation}
By a standard interpolation argument (see (\ref{Sobolevspectr})) the upper bound in (\ref{integers}) also holds for non-integer indices $s >1$. Actually, the same is true for the lower bound, since for any integer $n>s$ we have
\begin{align} \nonumber
\lbrace \Vert u\Vert_s^2 \rbrace &\geq \lbrace \Vert u\Vert_n^2 \rbrace^{n-s+1} \lbrace \Vert u\Vert_{n+1}^2 \rbrace^{-(n-s)} \overset{s}{\gtrsim} \nu^{-(2s-1)}.
\end{align}
\medskip \\ \indent
In all results in this section as well as in Section~\ref{turb}, the quantities estimated for a fixed trajectory of the noise, such as
$$
\max_{s \in [t,t+1],\ x \in S^1}{u^{\omega}_x}
$$
or maxima in time of Sobolev norms, can be replaced by their suprema over all smooth initial conditions (taken before considering the expected value). For instance, the quantity
$$
\E \max_{s \in [t,t+1]} |u^{\omega}(s)|^{\alpha}_{m,p}
$$
can be replaced by
$$
\E \sup_{u_0 \in C^{\infty}} \max_{s \in [t,t+1]} |u^{\omega}(s)|^{\alpha}_{m,p}.
$$
For the lower estimates, this is obvious. For the upper ones, this follows form the following pathwise version of Theorem~\ref{upperwmp}, and analogous pathwise versions of Theorem~\ref{uxpos} and of the upper estimates in Section~\ref{turb}. To prove these statements, it suffices to recast the original proofs in a pathwise setting (i.e., to work for a fixed $\omega$ instead of using the expected values).

\begin{theo} \label{pathwiseupper}
For $m \in \lbrace 0,1 \rbrace$ and $p \in [1,\infty]$, or for $m \geq 2$ and $p \in (1,\infty]$, there exist constants $\beta(m,p),m'(m,p)>0$ such that we have:
\begin{align} \nonumber
&\max_{s \in [t,t+1]} \left| u^{\omega}(s) \right|_{m,p} \overset{m,p}{\lesssim} (1+\max_{s \in [t-1,t+1]}{\Vert w^{\omega}(s) \Vert_{m'}})^{\beta} \nu^{-\gamma},
\\ \label{uniformineq}
&t \geq 2,\ \omega \in \Omega.
\end{align}
\end{theo}
\indent
On the other hand, in the results of this section and of Section~\ref{turb} the \textit{expected values} (and not the quantities themselves) can be replaced by their infima over all smooth initial conditions. For instance, the quantity
$$
\E \max_{s \in [t,t+1]} |u(s)|_{m,p}
$$
can be replaced by
$$
\inf_{u_0 \in C^{\infty}}  \E \max_{s \in [t,t+1]} |u(s)|_{m,p}.
$$

\section{Estimates for small-scale quantities} \label{turb}

In this section, we estimate small-scale quantities which characterise Burgulence in physical space (increments, flatness) as well as in Fourier space (energy spectrum). We fix $t$ satisfying $t \geq T_1$. Its precise value is not important, since all estimates in Section~\ref{sob} hold uniformly in $t$ provided that $t \geq T_1$ and the same is true for all estimates in this section. For the notation used here, see Subsection~\ref{agreeturb}.

\subsection{Results in physical space} \label{phys}

We begin by proving upper estimates for the functions $S_{p,\alpha}(\ell)$. In the proofs of the two following lemmas, constants denoted by $C$ depend only on $p,\alpha$.

\begin{lemm} \label{upperdiss}
For $\alpha \geq 0$ and $\ell \in [0,1]$,
$$
S_{p,\alpha}(\ell) \overset{p,\alpha}{\lesssim} \left\lbrace \begin{aligned} & \ell^{\alpha p},\ 0 \leq p \leq 1. \\ & \ell^{\alpha p} \nu^{-\alpha(p-1)},\ p \geq 1. \end{aligned} \right.
$$
\end{lemm}

\textbf{Proof.} 
We begin by considering the case $p \geq 1$. We have
\begin{align} \nonumber
S_{p,\alpha}(\ell) &= \Big\{ \Big( \int_{S^1}{|u(x+\ell)-u(x)|^p dx} \Big)^{\alpha} \Big\}
\\ \nonumber
& \leq \Big\{ \Big( \max_{x} |u(x+\ell)-u(x)|^{p-1} \int_{S^1}{|u(x+\ell)-u(x)| dx} \Big)^{\alpha} \Big\}.
\end{align}
By H{\"o}lder's inequality we get
\begin{align} \nonumber
S_{p,\alpha}(\ell) \leq & \Big\{ \Big( \int_{S^1}{|u(x+\ell)-u(x)| dx} \Big)^{\alpha p} \Big\}^{1/p} 
\\ \nonumber
& \times \Big\{ \max_{x} |u(x+\ell)-u(x)|^{\alpha p} \Big\}^{(p-1)/p}.
\end{align}
Since the space average of $u(x+\ell)-u(x)$ vanishes, we obtain that
\begin{align} \nonumber
S_{p,\alpha}(\ell) \leq & \Big\{ \Big(2 \int_{S^1}{(u(x+\ell)-u(x))^{+} dx} \Big)^{\alpha p} \Big\}^{1/p}
\\ \nonumber
& \times \Big\{ \max_{x} |u(x+\ell)-u(x)|^{\alpha p} \Big\}^{(p-1)/p}
\\ \label{difference}
\leq & C \ell^{\alpha} \Big\{ \max_{x} |u(x+\ell)-u(x)|^{\alpha p} \Big\}^{(p-1)/p},
\end{align}
where the second inequality follows from Theorem~\ref{uxpos}. Finally, by Theorem~\ref{avoir} we get
\begin{align} \nonumber
S_{p,\alpha}(\ell) & \leq C \ell^{\alpha} \Big\{ ( \ell |u|_{1,\infty} )^{\alpha p} \Big\}^{(p-1)/p} \leq
C \ell^{\alpha p} \nu^{-\alpha(p-1)}.
\end{align}
The case $p<1$ follows immediately from the case $p=1$ since now $S_{p,\alpha}(\ell) \leq S_{1,\alpha p}(\ell)$, by H{\"o}lder's inequality. $\qed$
\medskip
\\ \indent
For $\ell \in J_2 \cup J_3$, we have a better upper bound if $p \geq 1$.

\begin{lemm} \label{upperinert}
For $\alpha \geq 0$ and $\ell \in J_2 \cup J_3$,
$$
S_{p,\alpha}(\ell) \overset{p,\alpha}{\lesssim} \left\lbrace \begin{aligned} & \ell^{\alpha p},\ 0 \leq p \leq 1. \\ & \ell^{\alpha},\ p \geq 1. \end{aligned} \right.
$$
\end{lemm}

\textbf{Proof.} The calculations are almost the same as in the previous lemma. The only difference is that we use another upper bound for the right-hand side of (\ref{difference}). Namely, we have
\begin{align} \nonumber
S_{p,\alpha}(\ell) & \leq C \ell^{\alpha} \Big\{ \max_{x} |u(x+\ell)-u(x)|^{\alpha p} \Big\}^{(p-1)/p}
\\ \nonumber
& \leq C \ell^{\alpha} \Big\{ (2 |u|_{\infty} )^{\alpha p} \Big\}^{(p-1)/p} \leq C \ell^{\alpha},
\end{align}
where the third inequality follows from Theorem~\ref{avoir}. $\qed$
\smallskip \\ \indent

To prove lower estimates for $S_{p,\alpha}(\ell)$, we need a lemma. Loosely speaking, this lemma states that with a probability which is not too small, during a period of time which is not too small, several Sobolev norms are of the same order as their expected values. Note that in the following definition, (\ref{condi}-\ref{condii}) contain lower and upper estimates, while (\ref{condiii}) only contains an upper estimate.  The inequality $|u(s)|_{\infty} \leq  \max u_x(s)$ in (\ref{condi}) always holds, since $u(s)$ has zero mean value and the length of $S^1$ is $1$.

\begin{defi}
For a given solution $u(s)=u^{\omega}(s)$ and $K>1$, we denote by $L_K$ the set of all $(s,\omega) \in [t,t+T_0] \times \Omega$ such that
\begin{align} \label{condi}
&K^{-1} \leq |u(s)|_{\infty} \leq \max u_x(s) \leq K
\\ \label{condii}
&K^{-1} \nu^{-1} \leq |u(s)|_{1,\infty} \leq K \nu^{-1}
\\ \label{condiii}
&|u(s)|_{2,\infty} \leq K \nu^{-2}.
\end{align}
\end{defi}

\begin{lemm} \label{typical}
There exist constants $\tilde{C},K_1>0$ such that for all $K \geq K_1$, $\rho(L_K) \geq \tilde{C}$. Here, $\rho$ denotes the product measure of the Lebesgue measure and $\Pe$ on $[t,t+T_0] \times \Omega$.
\end{lemm}

\textbf{Proof.}
We denote by $A_{K}$, $B_{K}$ and $D_{K}$ the set of $(s,\omega)$ satisfying
$$
\text{"The upper estimates in (\ref{condi}-\ref{condiii}) hold for a given value of $K$"},
$$
$$
\text{"The lower estimates in (\ref{condi}-\ref{condii}) hold for a given value of $K$"}
$$
and
$$
\text{"The lower estimate in (\ref{condii}) holds for a given value of $K$"},
$$
respectively.
\\ \indent
Note that for $K \leq K'$, $L_K \subset L_{K'}$, and similarly for $A_K$, $B_K$ and $D_K$.
\\ \indent
By Lemma~\ref{GN} we get $|u|_{\infty} \geq C' |u|_{2,\infty}^{-1} |u|_{1,\infty}^2$ 
for some constant
\\
$C'>0$. Thus, for $\tilde{K} \geq \max(C',1) K^3$, we have $A_{K} \cap D_{K} \subset B_{\tilde{K}}$, and therefore:
$$
A_{K} \cap D_{K} \subset A_{\tilde{K}} \cap B_{\tilde{K}}=L_{\tilde{K}}.
$$
Consequently:
$$
\rho(L_{\tilde{K}}) \geq \rho(A_{K})+\rho(D_{K})-T_0.
$$
By Theorem~\ref{uxpos}, Theorem~\ref{avoir} and Chebyshev's inequality, the measure of the set $A_{\tilde{K}}$ tends to $T_0$ as $\tilde{K}$ tends to $+\infty$. So to prove the lemma's assertion, it remains to show that there exists $C>0$ such that for $K$ large enough we have $\rho(D_K) \geq C$. Using the upper estimate for $\lbrace |u|^2_{1,\infty} \rbrace$ in Theorem~\ref{avoir}, we get
$$
\lbrace |u|_{1,\infty} \One(|u|_{1,\infty} \geq K \nu^{-1}) \rbrace \leq C K^{-1} \nu^{-1}.
$$
Here, $\One(A)$ denotes the indicator function of an event $A$. On the other hand, we clearly have
$$
\lbrace |u|_{1,\infty} \One(|u|_{1,\infty} \leq K^{-1} \nu^{-1}) \rbrace \leq K^{-1} \nu^{-1}.
$$
Now, for $K_0>0$, consider the function
$$
g_{K_0}=|u|_{1,\infty} \One(K_0^{-1} \nu^{-1} \leq |u|_{1,\infty}  \leq K_0 \nu^{-1}).
$$
The lower estimate for $\lbrace |u|_{1,\infty} \rbrace$ in Theorem~\ref{avoir} and the relations above yield
$$
\lbrace g_{K_0} \rbrace \geq (C-CK_0^{-1}-K_0^{-1}) \nu^{-1} \geq C_0 \nu^{-1}
$$
for some constant $C_0$, uniformly for large enough values of $K_0$. Since $g_{K_0} \leq K_0 \nu^{-1}$, we get
$$
\rho(g_{K_0} \geq C_0 \nu^{-1}/2) \geq C_0 K_0^{-1} T_0/2.
$$
Since $g_{K_0} \leq |u|_{1,\infty}$, we obtain that
\begin{align} \nonumber
&\rho(|u|_{1,\infty} \geq C_0 \nu^{-1}/2) \geq C_0 K_0^{-1} T_0/2,
\end{align}
which implies the existence of $C'',K''>0$ such that $\rho(D_{K''}) \geq C''$ for $K \geq K''$. $\qed$

\begin{defi}
For a given solution $u(s)=u^{\omega}(s)$ and $K>1$, we denote by $O_K$ the set of all $(s,\omega) \in [t,t+T_0] \times \Omega$ such that the conditions (\ref{condi}), (\ref{condiii}) and
\begin{equation} \label{condii'}
K^{-1} \nu^{-1} \leq -\min u_x \leq K \nu^{-1}
\end{equation}
hold.
\end{defi}

\begin{cor} \label{typicalcor}
If $K \geq K_1$ and $\nu < K_1^{-2}$, then $\rho(O_K) \geq \tilde{C}$. Here, $\tilde{C}, K_1$ are the same as in the statement of Lemma~\ref{typical}.
\end{cor}

\textbf{Proof.} For $K = K_1$ and $\nu < K_1^{-2}$, the estimates (\ref{condi}-\ref{condii}) tell us that for $(s,\omega) \in L_K$,
$$
\max u_x(s) \leq K_1 < K_1^{-1} \nu^{-1} \leq |u_x(s)|_{\infty}.
$$
Thus, in this case we have $O_K=L_K$, and therefore
$$
\rho(O_K)=\rho(L_K) \geq \tilde{C}_0.
$$
Finally, we observe that since increasing $K$ while keeping $\nu$ constant increases the measure of $O_K$, the corollary's statement still holds for $K \geq K_1$ and $\nu < K_1^{-2}$. $\qed$
\medskip \\ \indent
Now we fix
\begin{equation} \label{K}
K=K_1,
\end{equation}
and choose
\begin{equation} \label{nu0eq}
\nu_0=\frac{1}{6} K^{-2};\ C_1=\frac{1}{4}K^{-2};\ C_2=\frac{1}{20}K^{-4}.
\end{equation}
In particular, we have $0<C_1 \nu_0 <C_2<1$: thus the intervals $J_i$ are non-empty and non-intersecting for all $\nu \in (0,\nu_0]$.

\begin{lemm} \label{lowerdiss}
For $\alpha \geq 0$ and $\ell \in J_1$,
$$
S_{p,\alpha}(\ell) \overset{p,\alpha}{\gtrsim} \left\lbrace \begin{aligned} &\ell^{\alpha p},\ 0 \leq p \leq 1. \\ & \ell^{\alpha p} \nu^{-\alpha(p-1)},\ p \geq 1. \end{aligned} \right.
$$
\end{lemm}

\textbf{Proof.}
By Corollary~\ref{typicalcor}, it suffices to prove that the inequalities hold uniformly for $(s,\omega) \in O_K$ with $S_{p,\alpha}(\ell)$ replaced by
\begin{equation} \nonumber
\Big( \int_{S^1}{|u(x+\ell)-u(x)|^p dx} \Big)^{\alpha}.
\end{equation}
For $\alpha \neq 1$, this fact follows from the case $\alpha=1$. Indeed, if for $(s,\omega) \in O_K$, we have
$$
\int_{S^1}{|u(x+\ell)-u(x)|^p dx} \overset{p}{\gtrsim} \ell^p\ (resp.\ \ell^p \nu^{-(p-1)}),
$$
then we also have
$$
\Big( \int_{S^1}{|u(x+\ell)-u(x)|^p dx} \Big)^{\alpha} \overset{p,\alpha}{\gtrsim} \ell^{\alpha p}\ (resp.\ \ell^{\alpha p} \nu^{-\alpha(p-1)}).
$$
Till the end of the proof we assume that
$$
(s,\omega) \in O_K.
$$ 
\\ \indent
\textbf{Case $\mathbf{p \geq 1,\ \alpha=1}$.} Denote by $z$ the leftmost point on $S^1$ (considered as $[0,1)$) such that $ u'(z) \leq - K^{-1} \nu^{-1}$. Since $|u|_{2,\infty} \leq K \nu^{-2}$, we have
\begin{equation} \label{smallux}
u'(y) \leq -\frac{1}{2} K^{-1}  \nu^{-1},\quad y \in [z-\frac{1}{2} K^{-2} \nu,z+\frac{1}{2} K^{-2} \nu].
\end{equation}
Since $\ell \leq C_1 \nu=\frac{1}{4}K^{-2} \nu$, by H{\"o}lder's inequality we get
\begin{align} \nonumber
\int_{S^1}&{|u(x+\ell)-u(x)|^p dx} \geq \int_{z-\frac{1}{4} K^{-2} \nu}^{z+\frac{1}{4} K^{-2} \nu}{|u(x+\ell)-u(x)|^p dx}
\\ \nonumber
&\geq (K^{-2} \nu/2)^{1-p} \Big( \int_{z-\frac{1}{4} K^{-2} \nu}^{z+\frac{1}{4} K^{-2} \nu}{|u(x+\ell)-u(x)| dx} \Big)^p
\\ \nonumber
&= C(p) \nu^{1-p} \Big( \int_{z-\frac{1}{4} K^{-2} \nu}^{z+\frac{1}{4} K^{-2} \nu}{ \Big(\int_{x}^{x+\ell}{- u'(y) dy } \Big) dx} \Big)^p
\\ \nonumber
&\geq C(p) \nu^{1-p} \Big( \int_{z-\frac{1}{4} K^{-2} \nu}^{z+\frac{1}{4} K^{-2} \nu}{ \frac{1}{2} \ell K^{-1} \nu^{-1} \ dx} \Big)^p = C(p) \nu^{1-p} \ell^p.
\end{align}
\\ \indent
\textbf{Case $\mathbf{p < 1,\ \alpha=1}$.} By H{\"o}lder's inequality we get
\begin{align} \nonumber
&\int_{S^1}{|u(x+\ell)-u(x)|^p dx} \geq \int_{S^1}{\Big((u(x+\ell)-u(x))^+\Big)^p dx}
\\ \nonumber
&\geq \Big( \int_{S^1}{\Big((u(x+\ell)-u(x))^+ \Big)^2 dx} \Big)^{p-1} \Big( \int_{S^1}{(u(x+\ell)-u(x))^+ dx} \Big)^{2-p}.
\end{align}
Using the upper estimate in (\ref{condi}) we get
\begin{align} \nonumber
&\int_{S^1}{|u(x+\ell)-u(x)|^p dx}
\\ \nonumber
&\geq \Big( \int_{S^1}{\ell^2 K^2 dx} \Big)
^{p-1} \Big( \int_{S^1}{(u(x+\ell)-u(x))^+ dx} \Big)^{2-p}.
\end{align}
Finally, since $\int_{S^1}{(u(\cdot+\ell)-u(\cdot))}=0$, we obtain that
\begin{align} \nonumber
\int_{S^1}{|u(x+\ell)-u(x)|^p dx}& \geq C(p) \ell^{2 (p-1)} \Big( \frac{1}{2} \int_{S^1}{|u(x+\ell)-u(x)| dx} \Big)^{2-p}
\\ \nonumber
&\geq C(p) \ell^p.
\end{align}
The last inequality follows from the case $p=1,\ \alpha=1$.\ \qed

\begin{rmq}
To prove this lemma, we do not need Corollary~\ref{typicalcor}. Indeed, in its proof we could have considered $z$ such that $|u'(z)| \geq K^{-1} \nu^{-1}$: Lemma~\ref{typical} guarantees its existence.
\end{rmq}

The proof of the following lemma uses an argument from \cite{AFLV}, which can be made rigorous if we restrict ourselves to the set $O_K$.

\begin{lemm} \label{lowerinert}
For $\alpha \geq 0$ and $\ell \in J_2$,
$$
S_{p,\alpha}(\ell) \overset{p,\alpha}{\gtrsim} \left\lbrace \begin{aligned} & \ell^{\alpha p},\ 0 \leq p \leq 1. \\ & \ell^{\alpha},\ p \geq 1. \end{aligned} \right.
$$
\end{lemm}

\textbf{Proof.}
For the same reason as in the previous proof, it suffices to prove that as long as $(s,\omega)$ belongs to $O_K$, the inequalities hold uniformly for $\alpha=1$ and for $S_{p,\alpha}(\ell)$ replaced by
$$
\int_{S^1}{|u(x+\ell)-u(x)|^p dx}.
$$
Once again, till the end of the proof we assume that $(s,\omega) \in O_K.$
\\ \indent
\textbf{Case $\mathbf{p \geq 1,\ \alpha=1}$.} Defining $z$ in the same way as previously, we have:
\begin{align} \nonumber
\int_{S^1}&{|u(x+\ell)-u(x)|^p dx} \geq
\\ \nonumber
&\int_{z-\frac{1}{2}\ell}^{z}
{ \Big| \int_{x}^{x+\ell}{u'^-(y)dy} - \int_{x}^{x+\ell}{u'^+(y)dy} \Big|^p dx}.
\end{align}
We have $\ell \geq C_1 \nu=\frac{1}{4} K^{-2} \nu$. Thus, by (\ref{smallux}), for $x \in [z-\frac{1}{2}\ell,z]$ we get
\begin{align} \nonumber
\int_{x}^{x+\ell}{u'^-(y)dy} &\geq \int_{z}^{z+\frac{1}{8} K^{-2} \nu}{u'^-(y) dy} \geq
\frac{1}{16} K^{-3}.
\\ \nonumber
&.
\end{align}
On the other hand, since $\ell \leq C_2$, using the upper estimate in (\ref{condi}) we get
$$
\int_{x}^{x+\ell}{u'^+(y)dy} \leq C_2 K \leq \frac{1}{20} K^{-3}.
$$
Thus,
\begin{align} \nonumber
& \int_{S^1}{|u(x+\ell)-u(x)|^p dx} \geq \frac{1}{2} \ell \Bigg( \Big(\frac{1}{16}-\frac{1}{20}\Big) K^{-3} \Bigg)^p \geq C(p) \ell.\ \qed
\end{align}
\\ \indent
\textbf{Case $\mathbf{p < 1,\ \alpha=1}$.} The result follows from the case $p=1,\ \alpha=1$ in exactly the same way as in the previous lemma.
\medskip \\ \indent
Summing up the results above we obtain the following theorem.

\begin{theo} \label{avoir2}
For $\alpha \geq 0$ and $\ell \in J_1$,
$$
S_{p,\alpha}(\ell) \overset{p,\alpha}{\sim} \left\lbrace \begin{aligned} & \ell^{\alpha p},\ 0 \leq p \leq 1. \\ & \ell^{\alpha p} \nu^{-\alpha (p-1)},\ p \geq 1. \end{aligned} \right.
$$
On the other hand, for $\alpha \geq 0$ and $\ell \in J_2$,
$$
S_{p,\alpha} (\ell) \overset{p,\alpha}{\sim} \left\lbrace \begin{aligned} & \ell^{\alpha p},\ 0 \leq p \leq 1. \\ & \ell^{\alpha},\ p \geq 1. \end{aligned} \right.
$$
\end{theo}

The following result follows immediately from the definition (\ref{flatness}).

\begin{cor} \label{flatnesscor}
For $\ell \in J_2$, the flatness satisfies $F(\ell) \sim \ell^{-1}$.
\end{cor}

\subsection{Results in Fourier space} \label{four}

By (\ref{integers}), for $m \geq 1$ we have
$$
\lbrace |\hat{u}^k|^2 \rbrace \leq (2 \pi k)^{-2m} \lbrace \Vert u \Vert_m^2 \rbrace \overset{m}{\sim} (k \nu)^{-2m} \nu.
$$
Thus, for $|k| \succeq \nu^{-1}$, $\lbrace|\hat{u}^k|^2\rbrace$ decreases super-algebraically.
\smallskip
\\ \indent
Now we want to estimate the $H^s$ norms of $u$ for $s \in (0,1)$.

\begin{lemm} \label{H12}
We have
$$
\lbrace \Vert u\Vert_{1/2}^2 \rbrace \sim |\log \nu|.
$$
\end{lemm}

\textbf{Proof.}
By (\ref{Sobolevfrac}) we have
\begin{align} \nonumber
\left\|u\right\|_{1/2} \sim \Bigg( \int_{S^1} \Big(\int_0^1 {\frac{|u(x+\ell)-u(x)|^2}{\ell^{2}} d \ell} \Big) dx \Bigg)^{1/2}.
\end{align}
Consequently, by Fubini's theorem,
\begin{align} \nonumber
\\ \nonumber
\lbrace \left\|u\right\|^2_{1/2} \rbrace &\sim \int_0^1 \frac{1}{\ell^2} \Big\{ \int_{S^1}{|u(x+\ell)-u(x)|^2 dx} \Big\} d \ell 
\\ \nonumber
&= \int_{0}^{1}{\frac{S_2(\ell)}{\ell^2} d\ell}
=\int_{J_1}{\frac{S_2(\ell)}{\ell^2} d\ell}+\int_{J_2}{\frac{S_2(\ell)}{\ell^2} d\ell}+\int_{J_3}{\frac{S_2(\ell)}{\ell^2} d\ell}.
\end{align}
By Theorem~\ref{avoir2} we get
$$
\int_{J_1}{\frac{S_2(\ell)}{\ell^2} d\ell} \sim \int_{0}^{C_1 \nu}{\frac{\ell^2 \nu^{-1}}{\ell^2} d\ell} \sim 1
$$
and
$$
\int_{J_2}{\frac{S_2(\ell)}{\ell^2} d\ell} \sim \int_{C_1 \nu}^{C_2}{\frac{\ell}{\ell^2} d\ell} \sim |\log  \nu|,
$$
respectively. Finally, by Lemma~\ref{upperinert} we get
$$
\int_{J_3}{\frac{S_2(\ell)}{\ell^2} d\ell} \leq C C_2^{-2} \leq C.
$$
Thus,
$$
\lbrace \left\|u\right\|^2_{1/2} \rbrace \sim |\log \nu|.\ \qed
$$
\smallskip \\ \indent
The proof of the following result follows the same lines.

\begin{lemm} \label{H01}
For $s \in (0,1/2)$,
$$
\lbrace \Vert u\Vert_s^2 \rbrace \overset{s}{\sim} 1.
$$
On the other hand, for $s \in (1/2,1)$,
$$
\lbrace \Vert u\Vert_s^2 \rbrace \overset{s}{\sim} \nu^{-(2s-1)}.
$$
\end{lemm}

The results above and the relation (\ref{integers}) tell us that $\lbrace|\hat{u}^k|^2\rbrace$ decreases very fast for $|k| \gtrsim \nu^{-1}$, and that for $s \geq 0$ the sums $\sum{|k|^{2s} \lbrace |\hat{u}^k|^2}\rbrace$ have exactly the same behaviour as the partial sums $\sum_{|k| \leq \nu^{-1}}{|k|^{2s} |k|^{-2}}$ in the limit $\nu \rightarrow 0^+$. Therefore we can conjecture that for $|k| \lesssim \nu^{-1}$, we have  $\lbrace|\hat{u}^k|^2 \rbrace \sim |k|^{-2}$.
\\ \indent
A result of this type actually holds (after layer-averaging), as long as $|k|$ is not too small. To prove it, we use a version of the Wiener-Khinchin theorem, stating that for any function $v \in L_2$ one has
\begin{equation} \label{spectrinertaux}
|v(\cdot+y)-v(\cdot)|^2=4\sum_{n \in \Z}{ \sin^2 (\pi ny) |\hat{v}^n|^2}.
\end{equation}
%

\begin{theo} \label{spectrinert}
If $M$ in the definition (\ref{spectrum}) of $E(k)$ is large enough, then for every $k$ such that $k^{-1} \in J_2$, we have $E(k) \sim k^{-2}$.
\end{theo}

\textbf{Proof.}
We recall that by definition,
$$
E(k) = \Bigg\{ \frac{\sum_{|n| \in [M^{-1}k,Mk]}{|\hat{u}^n|^2}}{\sum_{|n| \in [M^{-1}k,Mk]}{1}} \Bigg\}.
$$
Therefore proving the assertion of the theorem is the same as proving that
\begin{equation} \label{spectrinertequiv}
\sum_{|n| \in [M^{-1}k,Mk]}{ n^2 \lbrace |\hat{u}^n|^2 \rbrace } \sim k.
\end{equation}
The upper estimate is an immediate corollary of the upper estimate for $|u|_{1,1}$ in Theorem~\ref{avoir} and holds without averaging over $n$ such that $|n| \in [M^{-1} k, Mk]$. Indeed,  integrating by parts we get
\begin{equation} \nonumber
\lbrace |\hat{u}^n|^2 \rbrace \leq (2 \pi n)^{-2} \lbrace |u_x|^2_{1} \rbrace \leq C n^{-2},
\end{equation}
which proves the upper bound. Also, this inequality implies that
\begin{equation} \label{spectrinertupper1}
\sum_{|n| < M^{-1} k}{ n^2 \lbrace |\hat{u}^n|^2 \rbrace } \leq C M^{-1} k
\end{equation}
and
\begin{equation} \label{spectrinertupper2}
\sum_{|n| > M k}{\lbrace |\hat{u}^n|^2 \rbrace } \leq C M^{-1} k^{-1}.
\end{equation}
To prove the lower bound we note that
\begin{align} \nonumber
\sum_{|n| \leq M k}{ n^2 \lbrace |\hat{u}^n|^2 \rbrace } & \geq \frac{k^2}{\pi^2} \sum_{|n| \leq M k}{ \sin^2 (\pi nk^{-1}) \lbrace |\hat{u}^n|^2 \rbrace }
\\ \nonumber
&\geq \frac{k^2}{\pi^2} \Big( \sum_{n \in \Z}{ \sin^2 (\pi nk^{-1}) \lbrace |\hat{u}^n|^2 \rbrace } - \sum_{|n| > M k}{\lbrace |\hat{u}^n|^2 \rbrace } \Big).
\end{align}
Using (\ref{spectrinertaux}) and (\ref{spectrinertupper2}) we get
\begin{align} \nonumber
\sum_{|n| \leq M k}{ n^2 \lbrace |\hat{u}^n|^2 \rbrace } &\geq  \frac{k^2}{4 \pi^2} \Big( \lbrace |u(\cdot+k^{-1})-u(\cdot)|^2 \rbrace - C M^{-1} k^{-1} \Big)
\\ \nonumber
&\geq \frac{k^2}{4 \pi^2} (S_2(k^{-1})-C M^{-1} k^{-1}).
\end{align}
Finally, using Theorem~\ref{avoir2} we obtain that
\begin{equation} \nonumber
\sum_{|n| \leq M k}{ n^2 \lbrace |\hat{u}^n|^2 \rbrace }  \geq (C-C M^{-1}) k.
\end{equation}
Now we use (\ref{spectrinertupper1}) and we choose $M \geq 1$ large enough to obtain (\ref{spectrinertequiv}). $\qed$

\begin{rmq} \label{spectrinertrmq}
We actually have
$$
\Bigg\{ \Bigg( \frac{\sum_{|n| \in [M^{-1}k,Mk]}{|\hat{u}^n|^2}}{\sum_{|n| \in [M^{-1}k,Mk]}{1}} \Bigg)^{\alpha} \Bigg\} \overset{\alpha}{\sim} k^{-2\alpha},\quad \alpha>0.
$$
The upper bound is proved in the same way as above, and then the lower bound follows from H{\"o}lder's inequality and the lower bound in Theorem~\ref{spectrinert}.
\end{rmq}

\section{Stationary measure and related issues} \label{stat}

\subsection{A contraction property}
\indent
\medskip \\ \indent
Contraction properties for solutions of scalar conservation laws have been known to hold since the works of Oleinik and Kruzhkov (cf. \cite{Daf10} and references therein). In the space-periodic setting, we have the following contraction property in $L_1$.

\begin{theo} \label{contract}
Consider two solutions $u$, $\overline{u}$ of (\ref{whiteBurgers}), corresponding to the same realisation of the random force but different initial conditions $u_0,\overline{u}_0$ in $C^{\infty}$. For all $t \geq s \geq 0$, we have
\begin{equation} \nonumber
|u(t)-\overline{u}(t)|_{1} \leq |u(s)-\overline{u}(s)|_{1}.
\end{equation}
\end{theo}

\textbf{Proof.} We only consider the case $s=0$: the general case is proved in exactly the same way. Consider the function $v=u-\overline{u}$ and define
$$
\Phi(t,x)=\frac{f(u(t,x))-f(\overline{u}(t,x))}{u(t,x)-\overline{u}(t,x)}.
$$
Since $f$ is $C^{\infty}$-smooth and $u,\overline{u}$ are continuous in time and $C^{\infty}$-smooth in space, by Hadamard's lemma $\Phi$ is continuous in time and $C^{\infty}$-smooth in space. The function $v$ is a weak solution of the equation
\begin{equation} \label{direct}
v_t+( \Phi v )_x=\nu v_{xx},\ v(0)=v_0=u_0-\overline{u}_0,\ 0 \leq t \leq T.
\end{equation}
Moreover, since $u_t-w_t$ and $\overline{u}_t-w_t$ are $C^{\infty}$-smooth in space, the same is true for $v_t$. Consequently, $v$ is the classical solution of (\ref{direct}). Now we consider the dual parabolic problem
\begin{equation} \label{dual}
h_t+\Phi h_x=-\nu h_{xx},\ h(T,x)=h_T(x),\ 0 \leq t \leq T.
\end{equation}
For a $C^{\infty}$-smooth final condition $h_T$, this problem has a unique classical solution $h$, $C^1$-smooth in time and $C^{\infty}$-smooth in space \cite{Aub}. Integrating by parts in time and in space, we get
\begin{align} \nonumber
&\left\langle v(T), h_T \right\rangle - \left\langle v_0, h(0) \right\rangle = \int_0^T{ \left\langle v_t(t), h(t) \right\rangle + \left\langle v(t), h_t(t) \right\rangle\ dt}
\\ \nonumber
&= \int_0^T{ \left\langle -(\Phi(t) v(t))_x+\nu v_{xx}(t), h(t) \right\rangle\ dt}
\\ \label{intparts}
&+ \int_0^T{ \left\langle v(t), -\Phi(t) h_x(t)-\nu h_{xx}(t) \right\rangle\ dt}=0.
\end{align}
Now we choose a sequence of $C^{\infty}$-smooth functions $h^n_T,\ n \geq 0,$ which approximate $sgn(v(T))$ pointwise and satisfy $|h^n_T| \leq 1$. We consider the solution $h^n$ to the problem (\ref{dual}) for $h_T=h^n_T$. By the maximum principle  \cite{Lan}, we have $|h^n(t,x)| \leq 1$ for all $t \in [0,T]$, $x \in S^1$. Now we pass to the limit as $n \rightarrow \infty$. By (\ref{intparts}), we get:
$$
|v(T)|_1=\lim_{n \rightarrow \infty}{\left\langle v(T), h^n_T \right\rangle}=\lim_{n \rightarrow \infty}{\left\langle v_0, h^n(0) \right\rangle} \leq |v_0|_1.\ \square
$$

\subsection{Setting and definitions} \label{sett}

Since $C^{\infty}$ is dense in $L_1$, Theorem~\ref{contract} allows us to extend the stochastic flow corresponding to (\ref{whiteBurgers}) to the space $L_1$. Indeed, consider any $\F_0$-measurable $u_0 \in L_1$ and approximate it in $L_1$ by a sequence of smooth functions $u_{0n},\ n \geq 1$. Let $u_n^\omega(t)$ be the solutions to the equation (\ref{whiteBurgers}) with the corresponding initial data. By Theorem~\ref{contract}, for each $\omega$ the sequence $\{u_n^\omega(t)\}$ is fundamental in the space $C(0,T; L_1)$. Its limit $u^\omega(t)$
does not depend on the sequence $u_{0n}$. We will call this limit {\it the $L_1$-solution of (12)} corresponding to the initial condition $u_0$. It is straightforward that Theorem~\ref{contract} remains valid for $L_1$-solutions.
\\ \indent
By construction, for every $\omega$, $t \mapsto u^{\omega}(t,\cdot)$ is continuous in $L_1$, and solutions to (\ref{whiteBurgers}) are $L_1$-solutions. 
\\ \indent
Conversely, for any $T>0$, $L_1$-solutions  are solutions to (\ref{whiteBurgers}) for $t \geq T$. It suffices to prove this in the case of a deterministic initial condition $u_0$. We will use the following elementary lemma, inspired by \cite[Theorem 1.2.17.]{KuSh}.

\begin{lemm} \label{doubleCV}
Let X be a Banach space, and let $x_n \in X$ be a sequence converging to $x$. Assume that 
$f : X \rightarrow \R \cup \left\{ +\infty \right\}$ is a Borel functional such that $f_k: X \rightarrow \R,\ k \geq 1,$ 
is a sequence of bounded continuous functions converging to f pointwise, and
$$
f_k(x_n) \leq C,\quad k,n \geq 1.
$$
Then $f(x) \leq C$.
\end{lemm}

\textbf{Proof.} It suffices first to let $n \rightarrow \infty$, and then to let $k \rightarrow \infty$. $\qed$ 
\medskip \\ \indent
Now take $T_2>T_1>0$ and consider $\omega \in \Omega$, an initial condition $u_0 \in L_1$, and the corresponding smooth approximations $u_{0n},\ n \geq 1,$ as above. Let $u$ and $u_n,\ n \geq 1,$ be the corresponding $L_1$-solution (resp., solutions) to (\ref{whiteBurgers}).
Let $X$ be the space $C(T_1,T_2;L_1)$ and consider the functions $f_k=f \circ \pi_k$ with $\pi_k$ the Galerkin projections on the subspace spanned by $x \mapsto e^{ilx},\ |l| \leq k$, and $f$ the Borel functional
$$
v \mapsto \max_{s \in [T_1,T_2]} \left\|v(s)\right\|^2_m.
$$
We check that $f$ and the $f_k$ verify the assumptions of Lemma~\ref{doubleCV}. By Lemma~\ref{contract}, we have $u^{\omega}_n \rightarrow u^{\omega}$ in $X$. On the other hand, by a time-rescaled version of Lemma~\ref{pathwiseupper}, we know that there exist constants $\beta(m),m'(m)$ such that we have:
\begin{equation} \nonumber
f_k(u^{\omega}_n) \leq f(u^{\omega}_n) \overset{m,T_1,T_2}{\lesssim} (1+\max_{s \in [t-1,t+1]}{\Vert w^{\omega}(s) \Vert_{m'}})^{2\beta} \nu^{-(2m-1)},\quad k,n \geq 1.
\end{equation}
Now Lemma~\ref{doubleCV} yields
$$
f(u^{\omega}) \overset{m,T_1,T_2}{\lesssim} (1+\max_{s \in [t-1,t+1]}{\Vert w^{\omega}(s) \Vert_{m'}})^{2\beta} \nu^{-(2m-1)}.
$$
This proves that for every $\omega$, the $L_1$-solutions $u^{\omega}(t)$ are $C^{\infty}$-smooth for $t>0$. Moreover, for every $m \geq 0$ and $T_2>T_1>0$, the upper estimates in $H^m$ for those solutions are uniform with respect to $u_0$ and with respect to $t \in [T_1,T_2]$. By interpolation, we can prove that the $L_1$-solutions are limits of the corresponding approximations in every Sobolev space $H^m,\ m \geq 0$. This has two important implications:
\begin{itemize}
\item For any $T>0$, we can pass to the limit $n \rightarrow \infty$ in the relation (\ref{Burgersintbis}). This proves that the $L_1$-solutions $u(t)$ are solutions to (\ref{whiteBurgers}) for $t \geq T$.
\item We can extend the results of Sections~\ref{sob}-\ref{turb} to $L_1$-solutions.
\end{itemize}
\indent 
As in the case of smooth solutions, the $L_1$-solutions of (\ref{whiteBurgers}) form a continuous Markov process in the space $L_1$. So they define a Markov semigroup $S_t^{*}$, acting on Borel measures on $L_1$. Till the end of this section the $L_1$-solutions to (\ref{whiteBurgers}) will be referred to as {\it solutions}.
\\ \indent
A \textit{stationary measure} is a Borel probability measure on $L_1$ invariant by $S_t^{*}$ for every $t$.  A \textit{stationary solution} of (\ref{whiteBurgers}) is a random process $v$ defined for $(t,\omega) \in [0,+\infty) \times \Omega$, valued in $L_1$, which solves (\ref{whiteBurgers}), such that the distribution of $v(t,\cdot)$ does not depend on $t$. Such a distribution is automatically a stationary measure.
\\ \indent
Now we consider the question of existence and uniqueness of a stationary measure, which implies existence and uniqueness (in the sense of distributions) of a stationary solution. This fact has been proved in a slightly different setting: see \cite{IK} and references therein; see also \cite{EKMS} for the proof in the case $\nu=0$. Moreover, we obtain a bound for the rate of convergence to the stationary measure in an appropriate distance. This bound does not depend on the viscosity or on the initial condition.

\begin{defi}
Fix $p \in [1,\infty)$. For a continuous real-valued function $g$ on $L_p$, we define its Lipschitz norm as
$$
|g|_{L(p)}:=\sup_{L_p}{|g|}+|g|_{Lip},
$$
where $|g|_{Lip}$ is the Lipschitz constant of $g$. The set of continous functions with finite Lipschitz norm will be denoted by $L(p)=L(L_p)$. We will abbreviate $L(1)$ as $L$.
\end{defi}

\begin{defi}
For two Borel probability measures $\mu_1,\mu_2$ on $L_p$, we denote by $\Vert \mu_1-\mu_2 \Vert^*_{L(p)}$ the Lipschitz-dual distance:
$$
\Vert \mu_1-\mu_2 \Vert^*_{L(p)}:=\sup_{g \in L(p),\ |g|_{L(p)} \leq 1}{\Big| \int_{S^1}{g(v) \mu_1(dv)}-\int_{S^1}{g(v) \mu_2(dv)} \Big|}.
$$
\end{defi}

Existence of a stationary measure for (\ref{whiteBurgers}) can be proved using the Bogolyubov-Krylov argument (see \cite{KuSh}). Let us give a sketch of the proof.
\\ \indent
Let $u(s)$ be a solution of (\ref{whiteBurgers}). For $s \geq 1$, $\E |u(s)|_{1,1}$ is uniformly bounded. Since by Helly's selection principle \cite{KoFo75}, $W^{1,1}$ is compactly embedded in $L_1$, the family of measures $\mu_{t}$ defined by:
$$
\mu_{t}:=\frac{1}{t} \int_{1}^{1+t}{S_s^{*}\mu_{u_0}\ ds},\ t \geq 1,
$$
where $\mu_{u_0}$ denotes the measure on $L_1$ induced by an initial condition $u_0$, is tight in $L_1$ for any initial condition $u_0$. Thus, we can extract a subsequence $\mu_{t_n}$, converging weakly to a limit $\mu$. It is not hard to check that $\mu$ is a stationary measure for (\ref{whiteBurgers}) in $L_1$.
\\ \indent
The main result of this section is the following theorem, proved in Subsection~\ref{proof}.

\begin{theo} \label{algCV}
There exists a positive constant $C'$ such that we have
\begin{equation} \label{algCVformula}
\Vert S_t^{*}\mu_1-S_t^{*}\mu_2 \Vert^*_L \leq C't^{-1/13},\qquad t \geq 1,
\end{equation}
for any probability measures $\mu_1$, $\mu_2$ on $L_1$.
\end{theo}

\begin{cor} \label{algCVcor}
For every $p \in (1,\infty)$, there exists a positive constant $C'(p)$ such that we have
\begin{equation} \\
\Vert S_t^{*}\mu_1-S_t^{*}\mu_2 \Vert^*_{L(p)} \leq C't^{-1/13p},\qquad t \geq 1,
\end{equation}
for any probability measures $\mu_1$, $\mu_2$ on $L_p$.
\end{cor}
\indent
Corollary \ref{algCVcor} is proved similarly to Theorem~\ref{algCV}, observing that by H{\"o}lder's inequality, for any pair of solutions $u,\overline{u}$ of (\ref{whiteBurgers}) and $p \in [1,\infty)$ we have
\begin{align} \nonumber
|u-\overline{u}|_p & \lesssim (|u-\overline{u}|_1)^{1/p} (|u-\overline{u}|_{\infty})^{(p-1)/p}.
\end{align}
Note that all estimates in the previous sections still hold for a stationary solution, since they hold uniformly for any initial condition in $L_1$ for large times, and a stationary solution has time-independent statistical properties. It follows that those estimates still hold when averaging in time and in ensemble (denoted by $\lbrace \cdot \rbrace$) is replaced by averaging solely in ensemble, i.e. by integrating  with respect to $\mu$. In particular, Theorem~\ref{avoir}, Theorem~\ref{avoir2} and Theorem~\ref{spectrinert} imply, respectively, the following results.

\begin{theo}
For $m \in \lbrace 0,1 \rbrace$ and $p \in [1,\infty]$, or for $m \geq 2$ and $p \in (1,\infty]$,
\begin{equation} \nonumber
\Big( \int {\left|u\right|_{m,p}^{\alpha} \mu(du)} \Big)^{1/\alpha} \overset{m,p,\alpha}{\sim} \nu^{-\gamma},\quad \alpha>0.
\end{equation}
\end{theo}

\begin{theo}
For $\alpha \geq 0$ and $\ell \in J_1$,
$$
\int{ \Big( \int_{S^1}{|u(x+\ell)-u(x)|^p dx} \Big)^{\alpha} \mu(du)} \overset{p,\alpha}{\sim} \left\lbrace \begin{aligned} & \ell^{\alpha p},\ 0 \leq p \leq 1. \\ & \ell^{\alpha p} \nu^{-\alpha (p-1)},\ p \geq 1. \end{aligned} \right.
$$
On the other hand, for $\alpha \geq 0$ and $\ell \in J_2$,
$$
\int{ \Big( \int_{S^1}{|u(x+\ell)-u(x)|^p dx} \Big)^{\alpha} \mu(du)} \overset{p,\alpha}{\sim} \left\lbrace \begin{aligned} & \ell^{\alpha p},\ 0 \leq p \leq 1. \\ & \ell^{\alpha},\ p \geq 1. \end{aligned} \right.
$$
\end{theo}

\begin{theo}
For $k$ such that $k^{-1} \in J_2$, we have:
$$
\int{ \frac{\sum_{|n| \in [M^{-1}k,Mk]}{|\hat{u}^n|^2}}{\sum_{|n| \in [M^{-1}k,Mk]}{1}} \mu(du)} \sim k^{-2}.
$$
\end{theo}

\subsection{Proof of Theorem~\ref{algCV}} \label{proof}

To begin with, we need an auxiliary lemma. The main idea of the proof is similar to that of Theorem~\ref{uxpos}: namely, if the white noise is small during a certain time, then the solution itself becomes small. The technique is also similar: we apply the maximum principle to a well-chosen function. We only give the proof for an initial condition in $C^{\infty}$: the general case follows as above by considering smooth approximations.

\begin{lemm} \label{uxposgen}
There exists a constant $\tilde{C} \geq 2$ such that if $\tau \geq \tilde{C}$ and if for some $t \geq 0$ and $\omega \in \Omega$, the trajectory of the Wiener process $w^{\omega}$ satisfies
$$
K=\max_{s \in [t,t+\tau]}{|w^{\omega}(s)-w^{\omega}(t)|_{3,\infty}} \leq \tau^{-2},
$$
then the corresponding solution $u^{\omega}(t,x)$ to (\ref{whiteBurgers}) satisfies
\begin{equation} \label{uxposgenresult}
\max_{x \in S^1}{u_x(t+\tau,x)} \leq \tau^{-1/2}.
\end{equation}
\end{lemm}

In this subsection, from now on we denote by $C'$ various positive constants, independent of $\tilde{C}$.
\smallskip
\\ \indent
\textbf{Proof.} Assume the converse. We abbreviate $w(s)-w(t)$ as $\tilde{w}(s)$ and we use the notation
\begin{equation} \label{tildev}
\tilde{v}(s,x)=(s-t) (u_x(s,x)-\tilde{w}_x(s,x));\quad N=\max_{s \in [t,t+\tau],\ x \in S^1}{\tilde{v}(s,x)}.
\end{equation}
Since we assumed that (\ref{uxposgenresult}) does not hold, we have
\begin{equation} \label{uxposgenaux}
N>\tau(\tau^{-1/2}- K) > \tau^{1/2}/2.
\end{equation}
Now consider a point $(t_1,x_1)$ at which the maximum $N$ is achieved. In the same way as in the proof of Theorem~\ref{uxpos}, we show that at $(t_1,x_1)$ we have
\begin{equation} \label{maxpointbis}
f''(u) (\tilde{v}+(t_1-t)\tilde{w}_x)^2 \leq \tilde{v}- (t_1-t)^2 f'(u) \tilde{w}_{xx} + \nu (t_1-t)^2 \tilde{w}_{xxx}.
\end{equation}
On the other hand, by (\ref{poly}) (as in the proof of Theorem~\ref{uxpos}, we use the notation $\delta=2-h(1)$) we get
\begin{align} \nonumber
(t_1-t)^2 f'(u(t_1,x_1)) &\leq C' (t_1-t)^{2} \Big(1+|u(t_1,x_1)| \Big)^{2-\delta}
\\ \nonumber
&\leq C' (t_1-t)^{\delta} \Big((t_1-t)+ (t_1-t)|u(t_1,x_1)| \Big)^{2-\delta}
\\ \nonumber
&\leq C' \tau^{\delta} \Big( \tau^{2-\delta}+(N+\tau K)^{2-\delta} \Big),
\end{align}
since $(t_1-t)u$ is the zero space average primitive of $\tilde{v}+(t_1-t) \tilde{w}_x$. Thus we get
$$
\sigma (N-\tau K)^2 \leq N+C' K \tau^{\delta} (\tau^{2-\delta}+(N+\tau K)^{2-\delta})+K \tau^2.
$$
By assumption, we have $\tau \geq \tilde{C}$ and $K \leq \tau^{-2}$, and by (\ref{uxposgenaux}) we have $N >  \tau^{1/2}/2$. Therefore we have, on the one hand,
$$
\sigma (N-\tau K)^2 \geq C' N^2,
$$
and on the other hand,
$$
N+C' K \tau^{\delta} (\tau^{2-\delta}+(N+\tau K)^{2-\delta})+K \tau^2 \leq C' N^{2-\delta}.
$$
Thus, $N^{\delta} \leq C'$, and for $\tilde{C}$ large enough we have a contradiction with the fact that $N>\tau^{1/2}$.\ $\square$
\smallskip
\\ \indent
To prove the following theorem, we use the coupling method \cite[Chapter 3]{KuSh}. The situation is actually simpler than for the stochastic 2D Navier Stokes equation, which is the main subject of \cite{KuSh}. Indeed, in our setting the "damping time" needed to make the distance between two solutions small does not depend on the initial conditions, and by Theorem~\ref{contract} the flow of (\ref{whiteBurgers}) is $L_1$-contracting.
\medskip
\\ \indent
\textbf{Proof of Theorem~\ref{algCV}.} We can take $(\mu_1$, $\mu_2)=(\delta_{u_0},\delta_{\overline{u}_0})$; the general case follows by Fubini's theorem. Indeed, we have
\begin{align} \nonumber
&\Vert S_t^{*} \mu_1 -S_t^{*} \mu_2 \Vert^*_L = \sup_{g \in L,\ |g|_{L} \leq 1}{\Big| \int{g(v) S_t^{*}\mu_1(dv)} -\int{g(v) S_t^{*}\mu_2(dv)} \Big|}
\\ \nonumber
& \leq \sup_{g \in L,\ |g|_{L} \leq 1}{ \int{ \Big| \int{g(v) S_t^{*}\delta_{u_0}(dv)}-\int{g(v) S_t^{*}\delta_{\overline{u}_0}(dv)} \Big| \mu_1(du_0) \mu_2(d \overline{u}_0)} }
\\ \nonumber
& \leq \sup_{u_0 \in Supp\ \mu_1,\ \overline{u}_0 \in Supp\ \mu_2}{\Vert S_t^{*} \delta_{u_0} -S_t^{*} \delta_{\overline{u}_0} \Vert^*_L}.
\end{align}
Now we denote by $u(t), \overline{u}(t)$ the solutions of (\ref{whiteBurgers}) corresponding respectively to the initial conditions $u_0,\overline{u}_0$. By the definition of the Lipschitz-dual distance, we have 
\begin{align} \nonumber
\Vert S_t^{*} \delta_{u_0} -S_t^{*} \delta_{\overline{u}_0} \Vert^*_L & = \sup_{g \in L,\ \Vert g \Vert_{L} \leq 1}{\Big| \E\ g(u(t))-\E\ g(\overline{u}(t))\Big|}
\\ \nonumber
& \leq \E \sup_{g \in L,\ \Vert g \Vert_{L} \leq 1}{\Big| g(u(t))-g(\overline{u}(t))\Big|}
\\ \label{CVLp}
& \leq \E \Big( \min(2,|u(t)-\overline{u}(t)|_{1}) \Big).
\end{align}
To prove the theorem's statement, it suffices to obtain the inequality
\begin{align} \label{algaux}
&\Pe \Big(|u(n^{13})-\overline{u}(n^{13})|_{1} > \frac{2}{n} \Big) \leq \frac{\tilde{C}'}{n}
\end{align}
for large enough integers $n$. Indeed, this inequality yields that for large enough $t$ we have
\begin{align} \nonumber
&\E\ \Big( \min(2,|u(t^{13})-\overline{u}(t^{13})|_{1}) \Big)
\\ \nonumber
&\leq \E\ \Big( \min(2,|u(\left\lfloor  t \right\rfloor^{13})-\overline{u}(\left\lfloor  t \right\rfloor^{13})|_{1}) \Big)
\\ \nonumber
&\leq \frac{2}{\left\lfloor  t \right\rfloor} \Pe \Big(|u(\left\lfloor  t \right\rfloor^{13})-\overline{u}(\left\lfloor  t \right\rfloor^{13})|_{1} \leq \frac{2}{\left\lfloor  t \right\rfloor} \Big) 
\\ \nonumber
&+2 \Pe \Big(|u(\left\lfloor  t \right\rfloor^{13})-\overline{u}(\left\lfloor  t \right\rfloor^{13})|_{1} > \frac{2}{\left\lfloor  t \right\rfloor} \Big)
\\ \nonumber
& \leq \frac{2+2\tilde{C}'}{\left\lfloor  t \right\rfloor} \leq \frac{C'}{t},
\end{align}
Here, $\left\lfloor  t \right\rfloor$ denotes the integer part of $t$, and the first inequality follows from Theorem~\ref{contract}. 
\\ \indent
By Theorem~\ref{contract}, for every $n \geq 1$ we have
\begin{align} \nonumber
&\Pe \Big(|u(n^{13})-\overline{u}(n^{13})|_{1} > \frac{2}{n} \Big)
\\ \nonumber
& = \Pe \Big(\forall k \in [1,n^{11}]:\quad |u(kn^{2})-\overline{u}(kn^{2})|_{1} > \frac{2}{n} \Big).
\end{align}
Thus,
\begin{align} \nonumber
&\Pe \Big(|u(n^{13})-\overline{u}(n^{13})|_{1} > \frac{2}{n} \Big)
\\ \nonumber
& \leq \Pe \Big(\forall k \in [1,n^{11}]:\quad |u(kn^{2})|_{1} > \frac{1}{n}\quad or \quad |\overline{u}(kn^{2})|_{1} > \frac{1}{n} \Big)
\\ \nonumber
& \leq \Pe \Big(\forall k \in [1,n^{11}]:\quad \max_{x \in S^1} u_x(kn^{2}) > \frac{1}{n}\quad or \quad \max_{x \in S^1} \overline{u}_x(kn^{2}) > \frac{1}{n} \Big).
\end{align}
The second inequality holds since the functions $u(t,\cdot)$ and $\overline{u}(t,\cdot)$ have zero mean value. From Lemma~\ref{uxposgen}, it follows that for $n \geq \tilde{C}^{1/2}$ we can only have $\max_{x \in S^1} u_x(kn^{2}) > \frac{1}{n}$ or $\max_{x \in S^1} \overline{u}_x(kn^{2}) > \frac{1}{n}$ if
$$
\max_{t \in [(k-1)n^2,kn^2]} |w(t)-w((k-1)n^2)|_{3,\infty} > \frac{1}{n^4},
$$
and therefore we get:
\begin{align}  \nonumber
&\Pe \Big(|u(n^{13})-\overline{u}(n^{13})|_{1} > \frac{2}{n} \Big)
\\ \nonumber
&  \leq \Pe \Big(\forall k \in [1,n^{11}]:\quad \max_{t \in [(k-1)n^2,kn^2]} |w(t)-w((k-1)n^2)|_{3,\infty} > \frac{1}{n^4} \Big).
\end{align}
Since the increments of $w$ on the time intervals $[(k-1)n^2,kn^2]$ are independent, we get that for $n \geq \tilde{C}^{1/2}$:
\begin{align} \nonumber
&\Pe \Big(|u(n^{13})-\overline{u}(n^{13})|_{1} > \frac{2}{n} \Big)
\\ \nonumber
&  \leq \sideset{}{}\prod_{1 \leq k \leq n^{11}} \Pe \Big( \max_{t \in [(k-1)n^2,kn^2]} |w(t)-w((k-1)n^2)|_{3,\infty} > \frac{1}{n^4} \Big),
\end{align}
and then by the inequality (\ref{Gauss}) we get:
\begin{align}  \nonumber
&\Pe \Big(|u(n^{13})-\overline{u}(n^{13})|_{1} > \frac{2}{n} \Big) \leq \Bigg( \exp \Big(-\frac{n^{-8}}{2C'n^2} \Big) \Bigg)^{n^{11}} \leq e^{-C'n} \leq \frac{C'}{n}.\ \square
\end{align}

\section*{Acknowledgements}
\indent
I would like to thank my Ph.D. advisor S.Kuksin for formulation of the problem, advice and guidance. I am also very grateful to A.Biryuk, U.Frisch and K.Khanin for helpful discussions. Finally, I would like to thank all of the staff and faculty at the CMLS in Ecole Polytechnique for advice and support.

\end{document}